\documentclass[iop]{emulateapj}
\usepackage{apjfonts}
\pdfoutput=1

\usepackage[bookmarks=true, colorlinks=true, citecolor=blue,backref=true]{hyperref}

\newcommand{\mic}{\mu{\rm m}}

\newcommand{\galex}{{\it GALEX}}
\newcommand{\wise}{{\it WISE}}
\newcommand{\her}{{\it Herschel}}

\slugcomment{Version 2.0, 2018 Apr 13}
\journalinfo{Submitted to ApJ}

\shorttitle{DUST ATTENUATION CURVES}
\shortauthors{SALIM ET AL.}

\begin{document}

\title{Dust attenuation curves in 
 the local universe: demographics and new laws for star-forming galaxies and
 high-redshift analogs}

\author{Samir Salim\altaffilmark{1}, M\'ed\'eric
  Boquien\altaffilmark{2}}
\author{ Janice C.\ Lee\altaffilmark{3}}
\altaffiltext{1}{Department of Astronomy, Indiana University,
   Bloomington, IN 47404, USA} 
\altaffiltext{2}{Universidad de Antofagasta, Unidad de Astronom\'ia, 
  Antofagasta 1270300, Chile}
\altaffiltext{3}{Spitzer Science Center, Caltech,
   Pasadena, CA 91125, USA}
\email{salims@indiana.edu}

\begin{abstract}
  We study dust attenuation curves of 230,000 individual galaxies in
  the local universe, ranging from quiescent to intensely star-forming
  systems, using \galex, SDSS, and \wise\ photometry calibrated on
  {\it Herschel}-ATLAS. We use a new method of constraining SED fits
  with infrared luminosity (SED+LIR fitting), and parameterized
  attenuation curves determined with the CIGALE SED fitting code.
  Attenuation curve slopes and UV bump strengths are reasonably well
  constrained independently from one another. We find that
  $A_{\lambda}/A_V$ attenuation curves exhibit a very wide range of
  slopes that are on average as steep as the SMC curve slope. The
  slope is a strong function of optical opacity. Opaque galaxies have
  shallower curves--in agreement with recent radiate transfer
  models. The dependence of slopes on the opacity produces an apparent
  dependence on stellar mass: more massive galaxies having shallower
  slopes. Attenuation curves exhibit a wide range of UV bump
  amplitudes, from none to MW-like; with an average strength 1/3 of
  the MW bump. Notably, local analogs of high-redshift galaxies have
  an average curve that is somewhat steeper than the SMC curve, with a
  modest UV bump that can be to first order ignored, as its effect on
  the near-UV magnitude is 0.1 mag. Neither the slopes nor the
  strengths of the UV bump depend on gas-phase metallicity. Functional
  forms for attenuation laws are presented for normal star-forming
  galaxies, high-$z$ analogs and quiescent galaxies. We release the
  catalog of associated SFRs and stellar masses (GSWLC-2).
\end{abstract}


\section{Introduction} \label{sec:intro}

A dust attenuation curve (or a law) describes how a galaxy's
integrated luminosity arising from stellar and nebular continuum is
affected by internal dust at different wavelengths: from far
ultraviolet (UV), where the attenuation is usually the most severe, to
near infrared (IR), where it becomes mostly negligible. The knowledge
and the appropriate use of attenuation curves are critical for the
study of galaxy populations and, consequently, of galaxy evolution. In
particular, the knowledge of the attenuation curve is required for the
robust derivation of galaxy physical parameters, especially in the
absence of dust emission information from mid and/or far IR.  This is
often the case for high redshift galaxies
(e.g.,\citealt{bouwens14,smit14,bowler15,finkelstein15,oesch14}), and
will be common for early-universe galaxy populations to be studied by
JWST (unless accompanied by observations made by ALMA). Attenuation
curves serve as an input in galaxy simulations (e.g.,
\citealt{dave17}), constrain the physical properties of dust grains in
different environments, inform radiative transfer models, and
characterize dust-star distribution and geometry as a function of
galaxy properties.

Though a great deal of progress has been made in the study of galaxy
attenuation curves, there are many open questions
\citep{conroy13}. How diverse are attenuation curves from one galaxy
to another? Are there trends between curve properties and galaxy's
physical and geometrical properties, such as the mass, star formation
rate (SFR), or inclination?

That galaxies may exhibit a diversity of attenuation curves has its
roots in the investigation of {\it extinction} curves, which precedes
the work on attenuation curves by several decades. Extinction curves
along individual sightlines are most commonly determined using the
``pair method'' \citep{stebbins43}. Spectra or photometry of reddened
stars are compared to the spectra or photometry of unreddened stars of
the same spectral type. Extinction curves have been studied along
multiple lines of sight to individual stars in the Milky Way and Large
and Small Magellanic Clouds (LMC and SMC)
\citep{nandy75,nandy80,roccavolmerange81}, from which the average
total extinction curves currently in use for these galaxies have been
derived (e.g.,
\citealt{seaton79,cardelli89,fitzpatrick86lmc,prevot84,gordon03}). For
galaxies other than the MW and the Clouds, the extinction curves have
been studied only in special circumstances, such as in occulting pairs
(e.g., \citealt{white92,holwerda13,keel14}). Great diversity is seen
both between different lines of sight in a single galaxy and between
the average curves of MW, LMC and SMC. The diversity of curves can be
characterized by two main features: their slope in the UV/optical
wavelength range, and the presence or absence of additional absorption
at 1700-2700\AA\ (the near-UV range), known as the 2175-\AA\ bump, or
just the UV bump \citep{stecher65}. Extinction curves towards MW
sightlines show a range of slopes with various degrees of UV bump
strengths \citep{fitzpatrick86,fitzpatrick88,cardelli89}. Steeper
slopes tend to be associated with sightlines that do not penetrate
deeply into the molecular clouds. Averaging the curves for lines of
sight that pass only through diffuse dust, an average MW extinction
curve is derived that nevertheless has a relatively shallow (gray)
slope and quite pronounced UV bump. In contrast to the MW curve, the
SMC extinction curve is significantly steeper (power-law exponent
steeper by about 0.5) and most sightlines lack the UV bump
\citep{bromage83}. The LMC curve lies in between the MW and SMC curves
in terms of steepness, as well as by having an intermediate UV
bump. The drivers of these differences are currently not well known.

If attenuation curves were similar to extinction curves, one may
think, based on just three galaxies, that massive galaxies have
shallower and bumpier curves than lower mass galaxies. However, an
average extinction curve does not necessarily describe how the
integrated light of a galaxy is affected by the dust, i.e., their
attenuation curve. The latter includes, in addition to absorption and
scattering out of the line of sight, the scattering into the line of
sight and the effects of dust/star distribution and viewing
orientation (the ``geometry''; see review by \citealt{calzetti01},
also \citealt{cf00,conroy10,conroy13,chevallard13}). In addition to
possible variation as a function of galaxy mass, there is a question
of whether starbursting galaxies, and by extension, high-redshift
populations, differ from more normal star-forming galaxies at low
redshift.

There are two main methods for deriving attenuation curves: the
``comparison method'' and the ``model-based method''. They
fundamentally differ in the way in which they establish the level of
attenuation of galaxies. The comparison method does it empirically
(usually based on the nebular emission Balmer decrement, which serves
as a proxy for continuum attenuation), while the model-based method
uses attenuated stellar population synthesis models. Furthermore, the
comparison method produces an average curve for an aggregate of
galaxies, whereas the model-based method produces curves for
individual galaxies. Model-based methods started to be exploited
relatively recently.

The comparison method was first employed to local UV-selected
starburst galaxies \citep{calzetti94,calzetti00}. The motivation
behind focusing on starbursts, a population of galaxies that is
relatively rare in the local universe, was to provide guidance for the
interpretation of rest-frame UV emission of high-redshift ($z>1$)
populations (Lyman Break galaxies), which the local starburst
resemble. Furthermore, the {\it IUE} UV spectra \citep{kinney93} of
starbursts had better constraining power because of the higher signal
in UV spectra compared to the ``normal'' star-forming galaxies. The
slope of the Calzetti (``starburst'') attenuation curve is similar to
that of the MW extinction curve (when normalized by $A_V$, see Section
\ref{ssec:param}), i.e., it is shallow, but without the UV bump and
with somewhat less abrupt far-UV rise. This result forms the basis for
a picture in which local starburst galaxies and high-redshift
populations have shallow, bump-free (i.e., Calzetti) curves, whereas
more normal galaxies may or may not have either steeper, or bumpier
curves.

The attenuation curves of more normal SF galaxies have only recently
been derived, but with conflicting results. The pair-matching method
of \citet{wild11}, which like the comparison method uses the Balmer
decrement but does not explicitly require identification of
unattenuated galaxies, was employed to derive the attenuation curve of
a larger and more general population of local star-forming galaxies,
showing a diversity of attenuation curves as a function of physical
properties of galaxies, but yielding typical curves that were closer
to the Calzetti curve than the steep SMC
curve. \citet{battisti16,battisti17a} extend the Calzetti comparison
method to normal star-forming galaxies (using UV/optical photometry
instead of spectra), and finding almost identical (shallow, no bump)
curves. On the other hand, studies that use model-based methods tend
to find, on average, steeper slopes both locally
\citep{conroy10uv,leja17} and at higher redshift
\citep{arnouts13,kriek13,salmon16,reddy17}. Furthermore, many studies
present evidence for a moderate UV bump both in local normal
star-forming galaxies \citep{conroy10uv}, and to some extent in higher
redshifts ($z>1$) populations \citep{noll07,kriek13}.

In this paper we study the demographics of attenuation curves in the
local universe in a more extensive way, and further probe the issues
regarding the diversity of the curves, the presence of the bump and
the extent of differences between different populations. We also
present a discussion on why different methodologies may produce
systematically different results, especially in terms of curve
steepness.

We constrain individual dust attenuation curves for a very large
sample of galaxies, spanning those that are quiescent to intensely
star-forming.  To overcome some of the challenges involved with this
task we apply a novel version of energy-balance SED fitting method,
where the constraints from the IR dust emission are applied in a way
that is computationally practical and yet samples the parameter space
with high resolution. Our study builds on previous efforts to
constrain attenuation curves using model-based methods carried out on
local galaxies but on smaller scale \citep{burgarella05,leja17}, or
for larger samples but at higher redshifts
\citep{arnouts13,kriek13,salmon16,tress18}.

The current study also updates the \galex-SDSS-\wise\ Legacy Catalog
(GSWLC), a catalog of physical parameters of 700,000 galaxies in the
local universe from the Bayesian SED fitting \citep{s16}. The
robustness of GSWLC parameters was the result of the careful treatment
of input photometry, a wide range of physically motivated star
formation histories, the inclusion of emission line corrections, and
importantly, appropriate choices regarding the dust attenuation
curve. In this work we expand on those efforts by directly including
the IR constraints that further refine the SFRs. The sample and input
data are described in Section \ref{sec:sample}. Details regarding the
methodology of IR luminosity determination, the SED fitting and the
parameterization of attenuation curves is given in Section
\ref{sec:method}. The resulting attenuation curves, their dependence
on physical properties of galaxies, comparison with previous
attenuation and extinction curves, and functional fits are presented
in Section \ref{sec:results}. Section \ref{sec:disc} discusses our
results in a the context of previous studies, and Section
\ref{sec:conc} presents the summary of the results.

Throughout this work we assume a Chabrier IMF \citep{chabrier} and
WMAP7 flat cosmology ($H_0=70.4$ km s$^{-1}$ Mpc$^{-1}$,
$\Omega_m=0.272$).

\section{Samples and data} \label{sec:sample}

We utilize ultraviolet (UV) and optical photometry together with the
estimates of total IR luminosity to derive total dust attenuation
curves of a large number of individual galaxies. IR luminosities were
obtained from the mid-IR photometry (12 or 22 $\mic$), using
luminosity-dependent IR templates and further corrected using the
calibrations derived from a subset of galaxies that have far-IR
photometry. The resulting IR luminosities have an accuracy of
$\sim 0.1 dex$ which, together with the UV and optical photometry
allows the slopes of individual attenuation curves to determined with
a typical error of 0.25 in the power-law slope exponent, several times
smaller than the range of slope exponents.

Dust attenuation curves in this paper are determined for GSWLC-M
galaxies that have mid-IR and UV photometry. GSWLC sample construction
is described in detail in \citet{s16}. In short, GSWLC
contains all galaxies with SDSS DR10 redshifts below $z=0.3$, brighter
than $r_{\rm petro}=18.0$, and covered by far-UV (FUV) and near-UV
(NUV) observations from \galex\ (data release GR6/7). Because \galex\
observations span a wide range of depths, separate samples were
produced, and independent SED fitting was performed, for the shallow,
``all-sky'' (GSWLC-A), medium-deep (GSWLC-M) and deep (GSWLC-D) UV
surveys, which encompass 88, 49 and 7\% of SDSS target galaxies,
respectively. If a galaxy was covered by more than one UV survey it was
included in each of respective catalogs. Mid-IR observations (12 and
22 $\mic$) from \wise\ cover the entire sky. 

GSWLC-M balances UV depth and sample size, and so we base our {\it
  target sample} on it. GSWLC-M includes 358,121 galaxies, regardless
of a IR or UV detection. For the purposes of this paper we require a
detection in either 12 or 22 $\mic$ and in one UV band. When only one
UV band yields a detection it is invariably the NUV, which goes deeper
than the FUV. Requiring both NUV and FUV would reduce the sample size
but is not necessary because the slope, being parameterized, is
relatively well constrained by the overall behavior at longer
wavelengths. We verify that there are no systematic differences
between single and two-band results. There are 228,335 galaxies in our
target sample after IR and UV detection cuts, with the mean redshift of 0.10.

We will specially focus on $\sim 1/2$ of the sample classified as
{\it star-forming}, i.e., the galaxies that lie below the AGN demarcation
line of \citet{k03c} in the \citet{bpt} (BPT) emission-line
diagram. We find that galaxies can be securely classified with the BPT
diagram by requiring the S/N ratio of H$\alpha$ line to be greater than
10 and the S/N ratio of the other three lines to be greater than 2
. The usual S/N$>3$ cut is overly restrictive because the S/N ratio of line
ratios of closely separated line is actually higher than the S/N ratio
of individual lines \citep{juneau14}. 
More accurately, this selection removes galaxies with significant AGN
contribution (even if some may have relatively high levels of SF) and
galaxies with weak emission lines. The resulting ``star-forming''
sample contains 113,892 galaxies with sSFR$>-11$, i.e., contains most
of the galaxies that form the star-forming sequence (galaxy ``main
sequence'').

In this paper we also utilize the {\it calibration sample}, composed of
galaxies from \her\-ATLAS Data Release 1. \her-ATLAS is the most
extensive deep far-IR survey, covering 161 sq.\ deg corresponding to
three fields in GAMA spectroscopic survey \citep{valiante16}, which is
itself located within SDSS footprint. Specifically, we use catalog
version 1.2, from which we extract \her-ATLAS galaxies included in
SDSS DR10 spectroscopic survey (based on \citealt{bourne16} matching),
having $z<0.3$, $r_{\rm petro}<18.0$ (same cuts as GSWLC) and detected
at 22 $\mic$ with \wise. Furthermore, we require that calibration
galaxies be classified as star-forming (i.e., without significant AGN
based on the BPT diagram), in order to remove potential contamination
by non-stellar dust heating. The resulting \her-\wise-SDSS calibration
sample contains 1891 galaxies, and is a representative subsample of
the target sample, having the same distribution in sSFR$--M_*$ plane
and the same range of IR luminosities (Section \ref{ssec:lir}).

For the target sample we use UV photometry from \galex,
optical photometry from SDSS, and mid-IR photometry from \wise. We
apply various corrections to FUV and NUV photometry to correct for the
edge-of-the-detector effects and blending, as described in
\citet{s16}. Five-band SDSS photometry uses {\tt modelMag}
magnitudes. UV and optical fluxes are corrected for Galactic reddening
using prescriptions from \citet{peek13} and \citet{yuan13},
respectively.

For the target and calibration samples we use photometry from two
longer-wavelength \wise\ channels: W3 and W4, centered at 12 and 22
$\mu$m, respectively.  Photometry is taken from unWISE catalog
\citep{lang16}, which performed forced photometry on \wise\ images
using SDSS centroids and profiles as priors. For a sample that is a
mix of galaxies that are resolved and unresolved in \wise, the
prior-based photometry is more appropriate and less biased than the
PSF magnitudes from the official AllWISE Source Catalog.

Far-IR and sub-mm photometry from \her\, used for the calibration
sample, consists of 100 and 160 $\mu$m fluxes from PACS instrument,
and 250, 350 and 500 $\mu$m fluxes from SPIRE instrument, and is
described in \citet{valiante16}.

To derive gas metallicities, AGN strength and BPT classification we
utilize emission line fluxes and equivalent widths from the MPA/JHU
catalog. These data, based on SDSS DR7 spectroscopic reductions
\citep{tremonti04}, are available for 97\% of target sample galaxies.

\section{Methodology} \label{sec:method}

%
\subsection{Derivation of IR luminosities} \label{ssec:lir}

In this paper we use a novel approach to energy-balance SED
fitting. In short, instead of fitting an SED that includes both
stellar (UV/optical) and dust (IR) SEDs, we perform UV/optical SED
fitting that includes constraints on the dust emission directly from
the total IR luminosity, i.e., the IR luminosity is treated as a
``flux'' point. This approach, which we call SED+LIR fitting, enables
more robust results with larger number of model SEDs. Details of the
method are discussed in Section \ref{sec:sed}.

\begin{figure}
\epsscale{1.1} \plotone{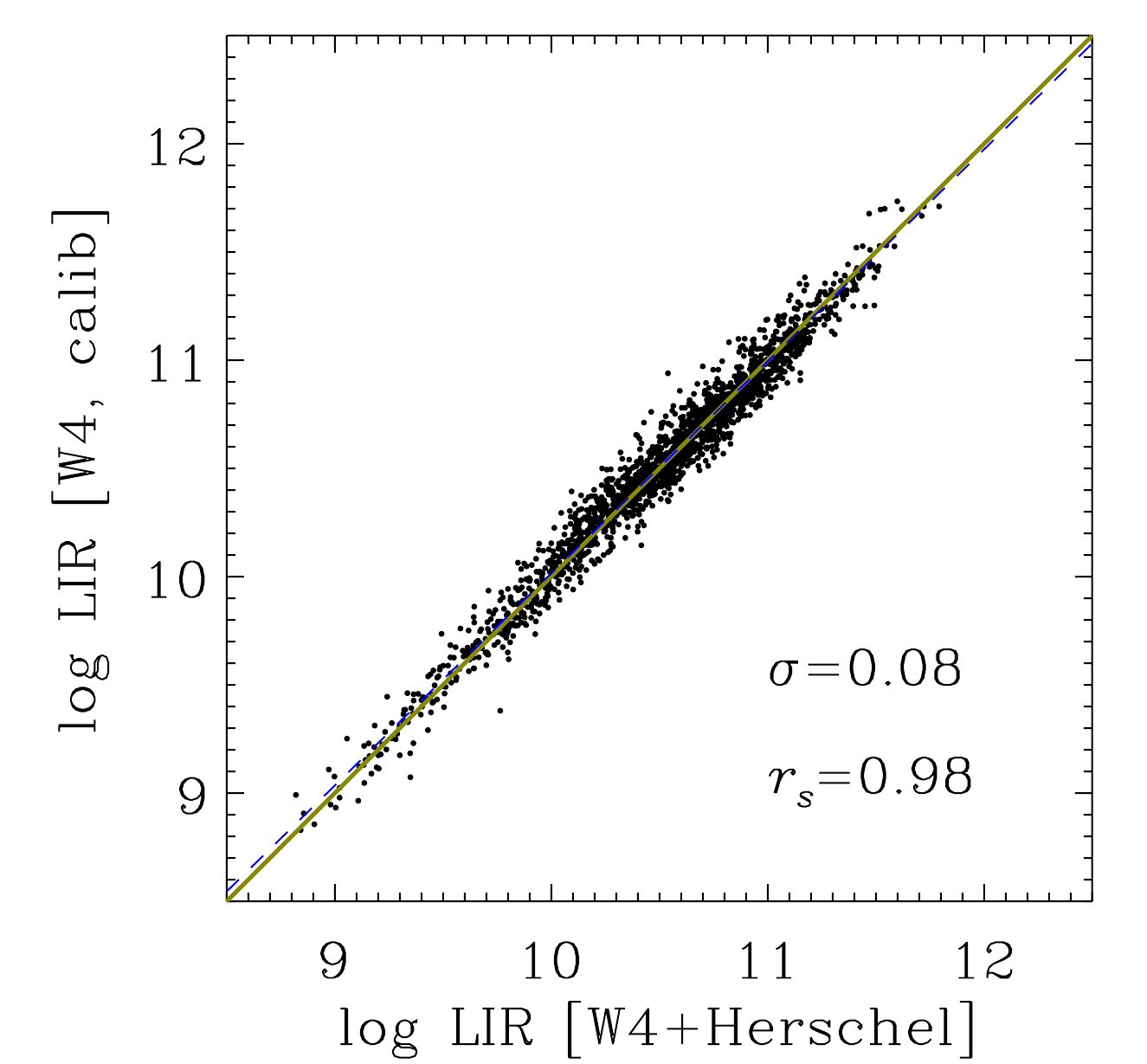}
\caption{Comparison between the IR luminosity derived from just the
  mid-IR flux point (22 $\mic$, \wise\ channel W4), using the new
  calibration described in the text, and the IR luminosity derived
  from the full IR SED from \her\ (PACS and SPIRE data in combination
  with W4). Our procedure is based on the use of luminosity-dependent
  templates plus the application of corrections, and results in
  estimates that have a small dispersion and no offset with respect to
  the full IR SED values. Green solid line is the 1:1 line, whereas
  the dashed blue line is the robust linear fit. Values of the
  standard deviation and the correlation coefficient are given in the
  plot. \label{fig:lir}}
\end{figure}

To obtain IR luminosities for our target sample, which typically
only has mid-IR photometry (from \wise), we perform a two-step
process. In Step 1 we derive an estimate for the IR luminosity based
on a single flux point (22 $\mic$, if detected, or 12 $\mic$
otherwise), using luminosity-dependent IR templates of
\citet{ce01}. For galaxies that belong to the \her\ calibration sample
we compare these estimates with the ``true'' IR luminosity obtained
from the full IR SED, and derive corrections
(calibration) to be applied to Step 1 estimates. True luminosities are
referred as such because they incorporate far-IR/sub-mm fluxes from
{\it Herschel} in addition to 22 $\mic$ flux from {\it WISE} and are 
determined by fitting \citet{ce01} templates (irrespective of the IR
luminosity associated with each template) to up to six flux
points (1 \wise, 2 PACS and 3 SPIRE) and using $\chi^2$ minimization
to find the best-fitting template and its scaling factor. The scaling
factor is applied to the luminosity of the template to arrive at the
final IR luminosity. True luminosities of the calibration sample span
three orders of magnitude $8.8<\log L_{\rm IR}<11.8$. Furthermore,
their distribution in the sSFR--$M_*$ diagram is essentially the same as
that of the full sample to which the calibration is applied, with the
average sSFRs being only 0.1 dex higher. In Step 2 we
apply the corrections to IR luminosities of galaxies in the target
sample. This method allows us to obtain accurate (Step 2) IR
luminosities without far-IR photometry. The accuracy is 0.08 (0.11)
dex for luminosities based on 22 (12) $\mic$ flux, with average
systematic offset smaller than 0.01 dex. The accuracy is similar
across all luminosities. The comparison between the IR
luminosities derived from just the 22 $\mic$ data and the full IR is
shown in Figure \ref{fig:lir}. The accuracy of our two-step method is
remarkable considering that the luminosities are ultimately based on a
single flux point in the mid-IR. Detailed description of the technique
and calibrations will be presented in a separate publication.

While most of the analysis in this paper focuses on galaxies selected
as star-forming, we perform the SED fitting and derive dust
attenuation parameters for non-star forming galaxies as well, which
includes the galaxies classified as AGN. The IR luminosities of AGN
may be affected by non-stellar dust heating, especially since they are
based on the mid-IR data. For galaxies classified as AGN, we find a
systematic trend between SFRs from GSWLC-1 (as we will refer to the
original version of the catalog), which were based solely on the
stellar emission, and the SFRs from a simple conversion from 22
$\mic$-based IR luminosity \citep{s16}. The trend increases with the
equivalent width of [OIII]5007 line, a proxy for AGN strength
\citep{k03c}. Based on this trend we derive a correction that is
applied to the IR luminosities of target sample galaxies classified as
AGN and is used in the SED fitting.

\subsection{Energy-balance SED fitting} \label{sec:sed}

This paper uses Code Investigating GALaxy Emission (CIGALE,
\citealt{noll09}, Boquien et al., in prep.) to perform the SED fitting and
determine physical properties of galaxies, such as the stellar mass
and the current SFR, but also, importantly, the dust attenuation
curve. In the construction of GSWLC-1 \citep{s16} we have performed
the SED fitting of only the UV and optical photometry, i.e., the
stellar and nebular emission. No constraints from the dust emission in
the IR were used. Instead, we have relied on the bulk comparison
between IR specific SFRs (sSFRs; derived using a simple conversion of
IR luminosities to SFR) and the SED fitting-derived sSFRs, to help us
select dust attenuation curves to use in the UV/optical SED
fitting. In particular, we have found that a good agreement between
UV/optical and IR sSFRs is achieved only if we assume an attenuation
curve that is significantly steeper than the one given in
\citet{calzetti00}, with further improvements in the level of
agreement when allowing the UV bump to be present in the curve (Fig.\
4 in \citealt{s16}).

In this work, IR luminosity is used in the SED fitting explicitly, by
applying a novel variant of {\it energy-balance} SED fitting.  IR
luminosity allows us to let the parameters that describe the
attenuation curve be free and determine then from the
fitting. Energy-balance SED fitting is the SED fitting that in
addition to stellar emission also takes into account dust emission in
the IR, such that the energy emitted by the dust (the IR luminosity)
matches the energy absorbed by the dust in the UV through near-IR. In
previous works, energy-balance SED fitting involved modeling both the
stellar (UV to near-IR) and dust (mid and far-IR) SEDs. Notably, such
approach is used in the SED fitting code MAGPHYS \citep{dacunha08},
where the IR SED is modeled as a composite of PAH template spectra,
the mid-IR continuum, and warm and cold grain emission
components. This model of IR SED requires six parameters, each with a
range of values, resulting in a large number of IR SEDs alone (50,000
in the case of MAGPHYS), each of which must be considered in
combination with a similar number of models describing the stellar
emission, leading to 2.5 billion model SEDs. Even if additional
constraints are used (e.g., that the sSFR and the IR SED shape are
correlated, \citealt{dacunha08}) the approach of simultaneously
modeling stellar and dust emission results in very large number of
parameter combinations. Similar approach was implemented in CIGALE,
where the specification of dust emission leads to a proliferation of
SED models that must be considered in the SED fitting. To keep the
energy-balance SED fitting computationally manageable while allowing
the parameters that describe the dust attenuation curve to be
unconstrained (which requires the number of models to be increased by
a factor of $\sim 50$--the number of different slope/bump
combinations), it would have been necessary to resort to coarse
sampling of the parameter space, potentially compromising the
precision of the derived parameters.

This paper takes another approach that allows us to keep the same
high-resolution model grid as the one used for just the UV/optical SED
fitting. We do so by decoupling the determination of the IR luminosity
from the SED fitting. Namely, we neither have the data nor need to fit
for the shape of the IR SED, since what constrains the SFR and the
attenuation curve is just the total IR luminosity. As shown in Section
\ref{ssec:lir}, by using luminosity-dependent IR templates together
with the empirical corrections, we can obtain accurate IR luminosities
from a single flux point. The key to our approach is to use the IR
luminosity as a {\it direct} constraint in the SED fitting, without
fitting the IR SED. We implement this approach using a custom
modification of CIGALE v0.11, whereby the observed IR luminosity is
treated as another SED ``flux'' point, i.e., the agreement between the
dust luminosity predicted by the model and the observed IR luminosity
goes into the $\chi^2$ that characterizes the quality of the fit.  In
this way we obtain all the benefits of the energy-balance SED fitting
without the need to proliferate the number of models, or be
potentially biased from the uncertainties stemming from the limited
coverage in the IR. We refer to this novel variant of energy-balance
SED fitting as IR luminosity-constrained SED fitting (SED+LIR
fitting).

Next, we provide an overview of other aspects of SED fitting, with
full details given in \citet{s16}. We use the same two-component
exponential model for star formation histories as used in GSWLC-1, a
parameterization consisting of an old population (formed 10 Gyr before
the present epoch), declining exponentially with varying $e$-folding
times, and a younger population (100 Myr to 5 Gyr old), having a mass
fraction between zero and 50\% of the old population, and a nearly
constant SFR. Such parameterization, yielding 1428 distinct SF
histories, overcomes the limitations of single-exponential decline
models (``$\tau$ models''), which, in order to produce high sSFRs
(blue colors) must be made artificially young and hence may miss the
mass from an old, fainter population (the ``outshining'' bias). The
models are calculated for four stellar metallicities (0.2 to 2.5
$Z_{\odot}$) using \citet{bc03} single stellar populations. The use of
\citet{m2005} models makes the resulting dust attenuation curves
steeper by 0.1. We use a Chabrier IMF, but confirm that the derived
values of dust attenuation parameters do not change beyond few percent
on average if Salpeter IMF is used instead.

An important feature of CIGALE is that it allows the calculation of
the contribution of emission lines to model broadband fluxes, the
omission of which leads to an additional 0.1 dex noise in SFR
determinations and 0.3 dex overestimate in the determination of SFRs
of high-sSFR galaxies (Fig.\ 5 in \citealt{s16}). Compared to GSWLC-1,
we now use new, improved emission line models with more extended range
and finer resolution of ionization parameters. The emission lines are
computed from the number of ionizing photons emitted by stellar
populations. The nebular templates expand upon \citet{inoue11} and
have been computed using Cloudy 08.00 \citep{ferland98}. They include
the nebular continuum (free-free, free-bound, and two-photon
processes) as well as 124 emission lines. The electron density is set
to 100 cm$^-3$. Furthermore, we now select both the ionization
parameter ($\log U=-3.4$) and the fraction of Lyman continuum photons
absorbed by dust ($f_{\rm dust}=0.3$), such that the model equivalent
widths of the main optical lines (H$\alpha$, H$\beta$, [OII], [OIII])
match, on average, the observed equivalent widths from SDSS
spectra. Our adopted fraction agrees with the estimates in the
literature \citep{inoue01,iglesias-paramo04}, although we find that
assuming a larger or smaller value has essentially no effect on the
results.

\subsection{Dust attenuation curves nomenclature} \label{ssec:nom}

Dust attenuation curves have been formulated in three different ways
in the literature, depending on the adopted normalization and whether
the curve is relative or not.  It is useful to review these
definitions and the corresponding nomenclature. We refer to the three
formulations as the {\it selective}, {\it total} and {\it absolute}
attenuation curves, which in their most fundamental form can be
expressed as:
$$
{\rm Selective}:\qquad \frac{A_{\lambda}-A_V}{A_B-A_V}
$$
$$
{\rm Total}:\qquad \frac{A_{\lambda}}{A_B-A_V}
$$
$$
{\rm Absolute}:\qquad \frac{A_{\lambda}}{A_V}
$$
\noindent where the attenuation of the continuum at some wavelength,
$A_{\lambda}$, is directly related to the optical depth. Two galaxies
can have the same attenuation law with one having higher $A_{\lambda}$
than the other, meaning it is dustier. Thus, in order to obtain the
attenuation curve, $A_{\lambda}$ must be normalized. In selective and
total formulations the normalization is done by color excess
$E(B-V)\equiv A_B-A_V$. The {\it total} formulation of the attenuation
curve is often denoted as $k_{\lambda}$:

\begin{equation}
k_{\lambda} = \frac{A_{\lambda}}{E(B-V)} = \frac{A_{\lambda}}{A_B-A_V}.
\end{equation}

\noindent The value of the total attenuation curve in $V$ band is
$k_V = A_V/E(B-V) \equiv R_V$, and is known as the ratio of total to
selective extinction in $V$. For Milky Way and LMC extinction curves
it is customary to use $R_V=3.1$, an average of different sightlines
through the diffuse ISM, having a range of values between 2 and 6
\citep{cardelli89}. The Calzetti curve is associated with
$R_{V, {\rm Cal}}=4.05$, with a galaxy-to-galaxy scatter of 0.8
\citep{calzetti00}. The {\it selective} formulation has mostly been
used in the older literature, where the total curve could not be
obtained because of the lack of IR data needed to anchor the curve.
The selective curve (sometimes designated as $E(\lambda-V)/E(B-V)$) is
related to the total curve as $k_{\lambda} -R_V$, i.e., it only gives
the relative attenuation with respect to $V$.\footnote{The selective
  attenuation curve is what \citet{reddy15} refer to as $fQ(\lambda)$,
  with $Q(\lambda)$ being the non-normalized selective curve
  introduced in \citet{calzetti94}.}.

 The third way to formulate the attenuation curve is to normalize
$A_{\lambda}$ by attenuation in $A_V$. This form is related
to the total curve formulation through:

\begin{equation}
A_{\lambda}/A_V = k_{\lambda}/k_V = k_{\lambda}/R_V.
\end{equation}

\noindent Normalization by absolute attenuation ($A_V$) is more
intuitive (and more fundamental, \citealt{cardelli89}) than by color
excess because the steepness of the attenuation curve in different
parts of the spectrum has direct interpretation. (In contrast, in
the total formulation the slope of the curve between $B$ and $V$ is
identical by definition.)  Furthermore, it is this form that is
required and sufficient to model the dust attenuation of galaxy
spectra. For historical reasons the normalization of the curve is tied
to $V$ band, though ideally it should be at longer wavelengths,
because different sightlines in our galaxy still show some variations
in extinction curves around $V$ \citep{cardelli88}. This formulation,
which, following \citet{cardelli88} we refer to as the {\it absolute},
will be used to discuss the results in this paper. Thus, whenever we
refer to a curve being steeper than another, we will mean steeper in
$A_{\lambda}/A_V$.

Note that two curves that have the same slopes in total formulation,
but are offset from one another (i.e., have different $R_V$) will not
have the same slopes in absolute formulation (and thus will not affect
the light in the same way). Lower values of $R_V$ mean that for the
same $A_V$ the attenuation at $B$ would be higher, making the {\it
  absolute} attenuation curve steeper in the optical range (but, as it
happens, also in the UV, \citealt{cardelli88}). Conversely, two curves
that have the same slope in absolute formulation (i.e., they are the
same curve, since the value of absolute curve at $V$ is always 1),
will not correspond to the same curves in total formulation unless
their $R_V$ values are the same.

\subsection{Parameterization of dust attenuation curves} \label{ssec:param}

With just the broadband photometry one cannot derive the attenuation
curve in detail, which is why it is useful to parameterize it.  Hence,
we follow the methodology of \citet{noll09}, implemented in CIGALE,
whereby the attenuation curve is defined as a two-parameter
modification of the total formulation of the Calzetti curve given in
Eq.\ 4 of \citet{calzetti00}\footnote{For the far-UV region (912-1500
  \AA), the Calzetti curve (defined at $\lambda>1200$ \AA) is
  substituted with the curve from \citet{leitherer02}. For simplicity,
  we will refer to this composite as the just the ``Calzetti curve''.}
The exact implementation differs to some extent from that in
\citet{noll09}, so we provide an updated overview here.

\begin{figure}
\epsscale{1.2} \plotone{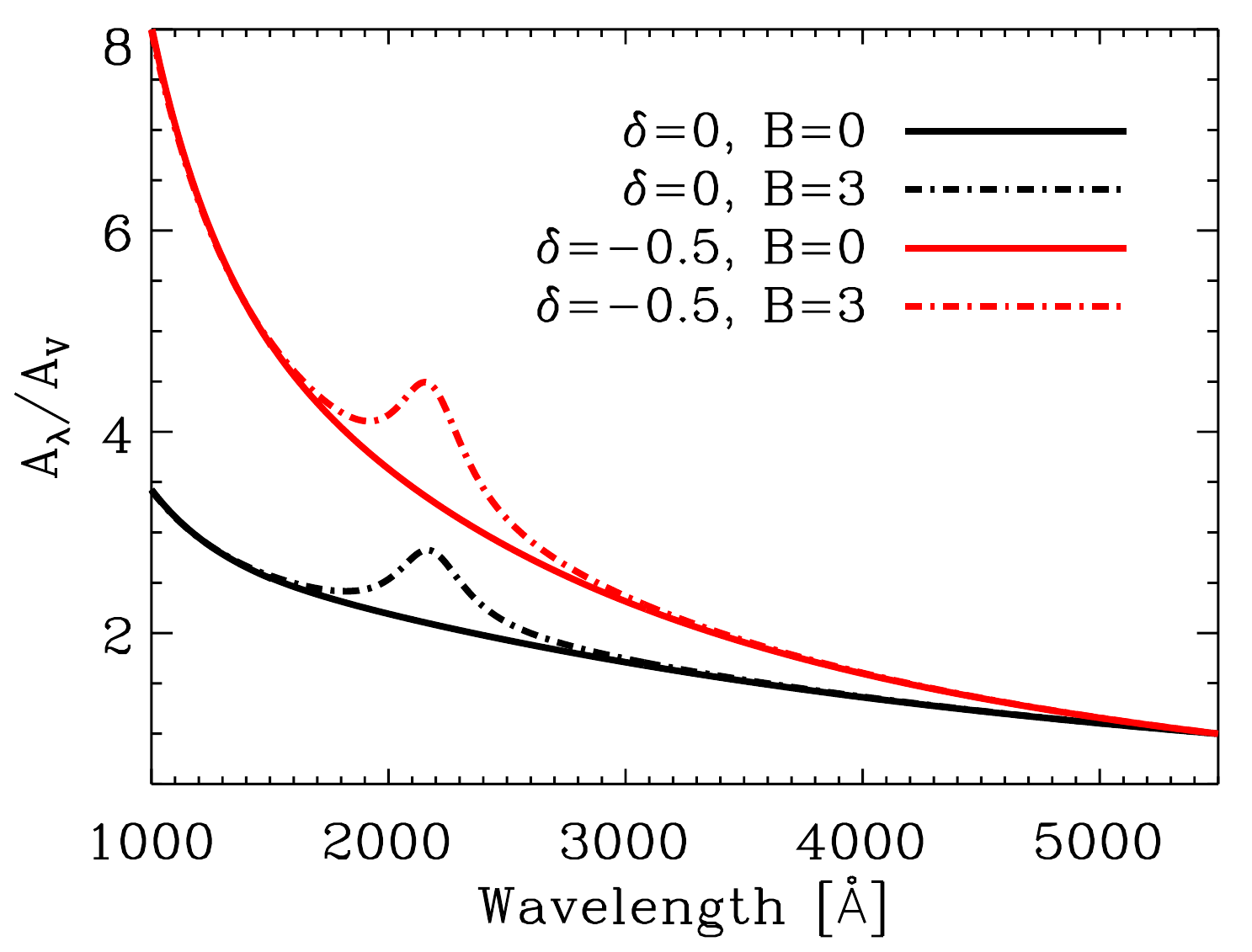}
\caption{Parameterization of dust attenuation curves. Following
  \citet{noll09}, we define the dust attenuation curve as a
  two-parameter modification of the Calzetti curve (black solid line),
  shown in absolute formulation ($A_{\lambda}/A_V$). Parameter
  $\delta$ modifies the power-law slope of the curve (with negative values
  making it steeper in the UV/optical region), while $B$ specifies the
  amplitude of the 2175\AA bump in the total formulation of the curve
  ($A_{\rm bump}/E(B-V)$). Fixed $B$ corresponds to roughly similar
  level of contribution of the attenuation due to the bump
  ($A_{\rm bump}$) to the total attenuation at 2175 \AA\
  ($A_{2175}$). \label{fig:scheme}}
\end{figure}

The first modification consists of allowing the slope of the curve to
deviate from the slope of the Calzetti curve. This is achieved by
multiplying the Calzetti curve ($k_{\lambda, {\rm Cal}}$) with a power
law term having an exponent $\delta$, ``centered'' at
$V$-band. Negative values of slope deviation $\delta$ produce
attenuation curves that are steeper in $A_{\lambda}/A_V$ than the
Calzetti curve at $\lambda<5500$ \AA\ (Figure \ref{fig:scheme}). The
Calzetti curve has $\delta=0$ by definition.

The second modification consists of adding a UV bump, following a
Drude profile $D_{\lambda}$ \citep{fitzpatrick86} given in Eq.\
\ref{eqn:drude}.\footnote{\citet{noll09} bump profile is equivalent to
  the one given by \citet{fitzpatrick86}, but the first uses $\gamma$
  to denote the width in units of wavelength, while in the other
  $\gamma$ denotes width in units of {\it inverse} wavelength.} The
strength of the bump is specified by amplitude $B$, in units of
$A_{\rm bump}/E(B-V)$, i.e., it pertains to the total formulation of
attenuation curve, where $A_{\rm bump}$ is the extra attenuation at
2175 \AA\ due to the bump. We keep the central wavelength and the
width of the bump fixed at default CIGALE values of 2175 \AA\ and 350
\AA, respectively.  The \citet{cardelli89} Milky Way curve has a bump
with a value of $B_{\rm MW}=3$. It needs to be pointed out that since
$B$ is defined in total curve formulation, it does not automatically
give the level of contribution of the attenuation due to the bump
($A_{\rm bump}$) to the total attenuation at 2175 \AA\
($A_{2175}$). However, it happens that modified Calzetti curves have
similar $k_{2175}=A_{2175}/E(B-V)$ regardless of $\delta$, which
implies that $A_{\rm bump}/A_{2175}\propto B$. Furthermore, since the
Calzetti curve and the MW extinction curve have similar slopes in
absolute formulation ($A_{\lambda}/A_V$), but different values of
$R_V$, it follows that for the same value of $B_{\rm MW}=3$, the
contribution of such bump to 2175 \AA\ attenuation in the Calzetti
curve will be $R_{V,{\rm MW}}/R_{V,{\rm Cal}}\sim 3/4$ of the
contribution in the Milky Way curve.
 
To summarize, the modified Calzetti attenuation law is given as:

\begin{equation}
k_{\lambda,{\rm mod}} = k_{\lambda,{\rm Cal}}\,\frac{R_{V,{\rm mod}}}{R_{V,{\rm Cal}}}
\left (\frac{\lambda}{5500\text{\AA}}\right)^{\delta}+D_{\lambda}
\label{eqn:mod1}
\end{equation}

\noindent where $R_{V,{\rm mod}}$ is the ratio of total to selective
extinction for the modified law, and is thus dependent on $\delta$,
while $R_{V,{\rm Cal}}=4.05$.\footnote{The implementation of
  the modified Calzetti law in CIGALE v0.11 and previous versions of
  the code adds the bump {\it before} applying the power-law
  modification. However, doing so effectively lowers the bump strength
  for a given input amplitude, leading to derived values of $B$
  becoming inflated by a factor $R_{V,{\rm Cal}} /R_{V,{\rm
      mod}}$. Our modification of the code implements the bump
  according to Eq.\ \ref{eqn:mod1}. This new method will be the
  default in CIGALE v0.12 and later.} The relationship between
$R_{V,{\rm mod}}$ and $\delta$ can be derived by imposing $E(B-V)=1$:

\begin{equation}
R_{V,{\rm mod}} = \frac{R_{V,{\rm Cal}}}{(R_{V,{\rm
      Cal}}+1)(4400/5500)^{\delta}-R_{V,{\rm Cal}}}.
\label{eqn:mod2}
\end{equation}

\noindent For example, for $\delta=-0.5$, $R_{V,{\rm mod}}=2.54$.

\begin{figure*}
\epsscale{1.15} \plotone{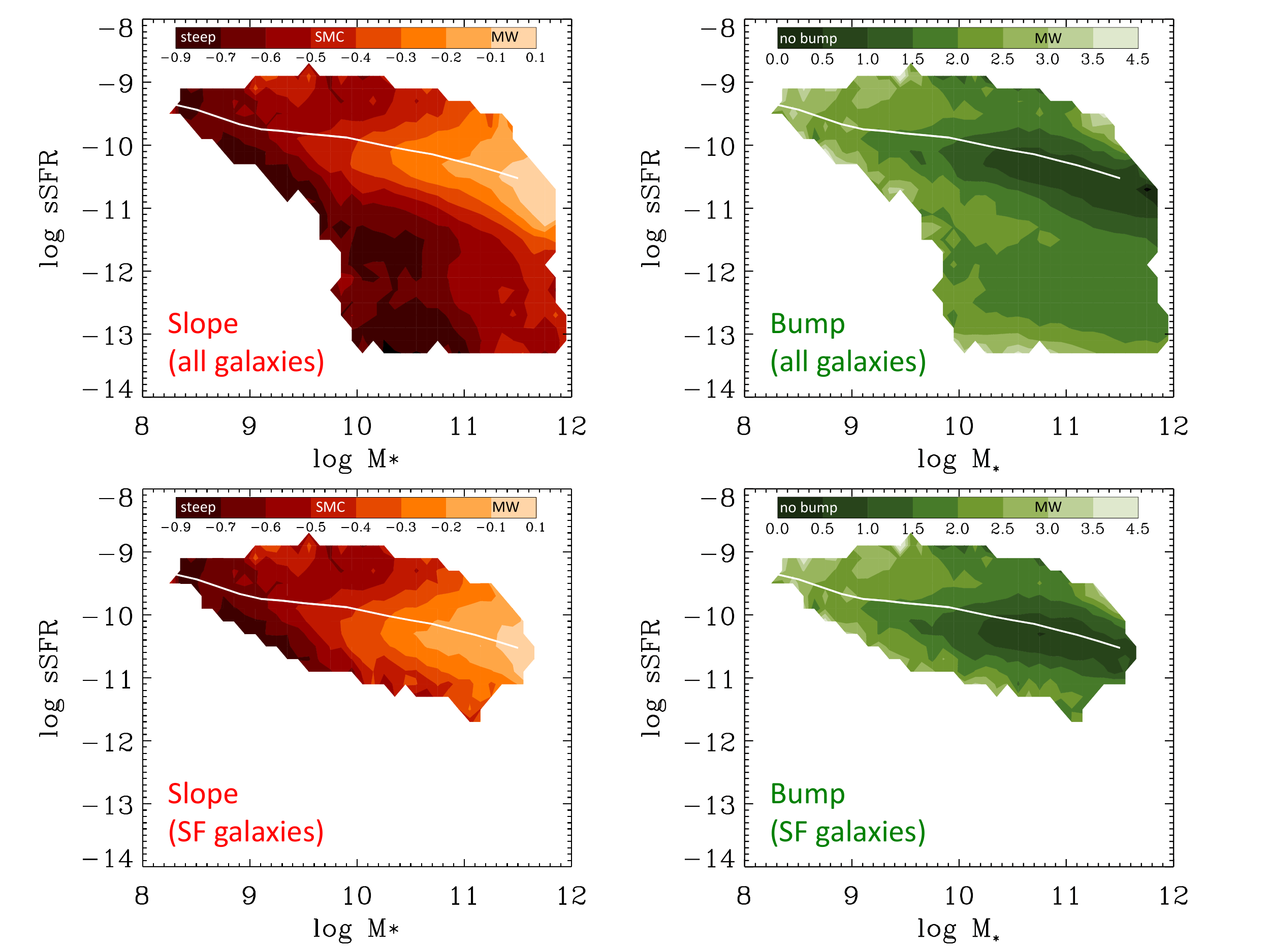}
\caption{Contour maps of dust attenuation curve parameters (average
  slope (left panels) and average UV bump strength (right panels)) as
  a function of the specific SFR and stellar mass, for all galaxies
  (upper panels) and galaxies classified as star-forming (lower
  panels). White line represents the median sSFR for star-forming
  galaxies (the ``main sequence'' of SF). The slope is expressed as
  the exponent of the power-law deviation ($\delta$) with respect to
  the Calzetti curve, the latter being represented by the lightest
  contour ($\delta=0$). There is a wide range of curve slopes, with
  lower-mass galaxies having steeper slopes. Bump strengths
  (amplitudes in units of $A_{\rm bump}/E(B-V)$, which for the Milky
  Way extinction curve has a typical value of 3) are on average weaker
  than the MW extinction curve bump, especially for star-forming
  galaxies of higher mass. Bins with 10 or more galaxies are
  shown. \label{fig:ssfrm_map}}
\end{figure*}

\begin{figure*}
\epsscale{1.15} \plotone{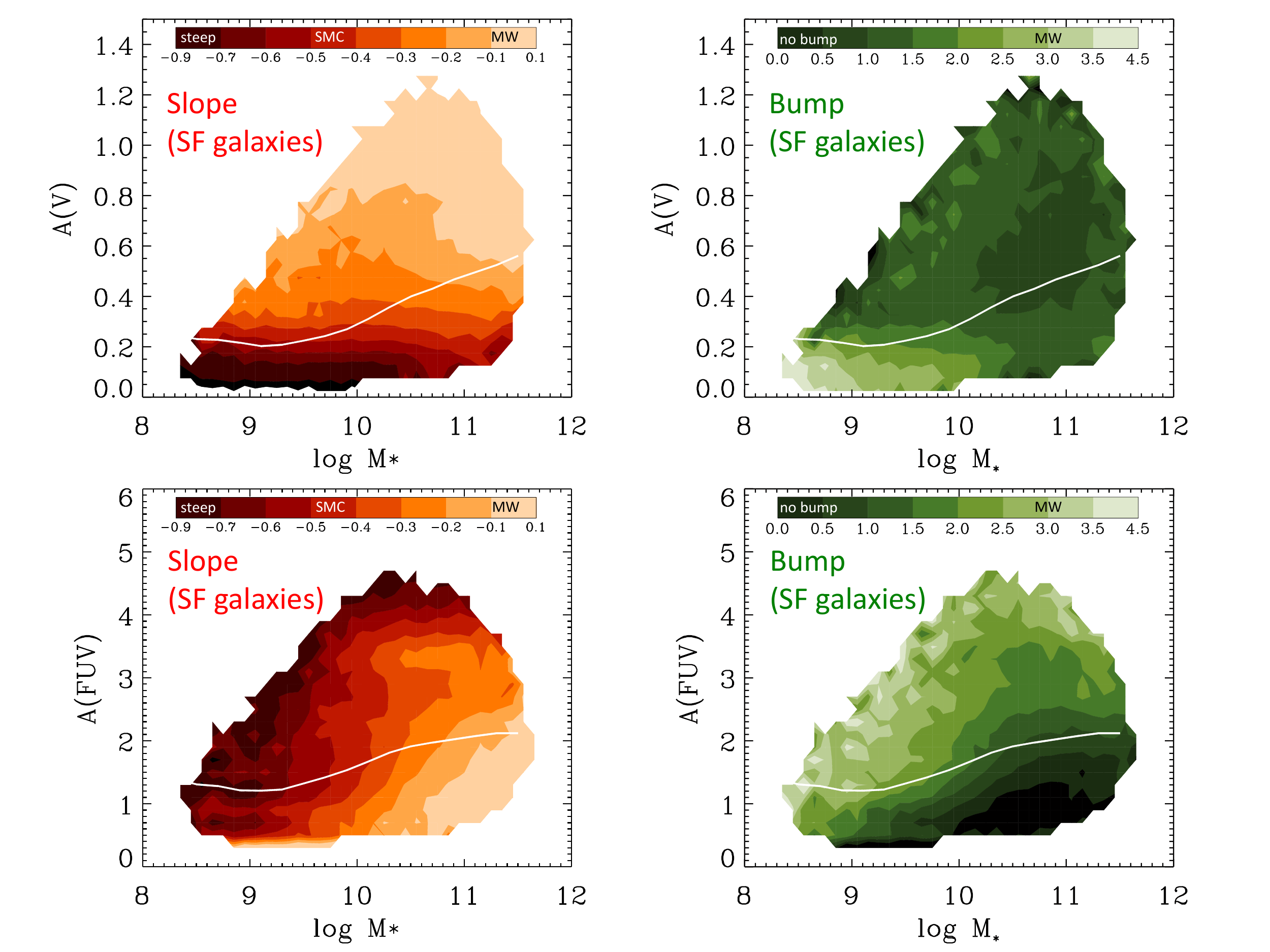}
\caption{Contour maps of dust attenuation curve parameters as a
  function of stellar mass and the level of dust attenuation, in $V$
  (upper panels) and FUV (lower panels). White lines are median
  trends. Attenuations are effective. The slope is strongly correlated
  with $A_V$, being steeper for low $A_V$ (and consequently higher
  $A_{\rm FUV}$). The UV bump is stronger for galaxies with lower
  optical opacity.\label{fig:atten_map}}
\end{figure*}

We emphasize that in this work we will be deriving attenuation curves
that affects individual (young and old) stellar populations. Such
curves are more physically relevant than the effective attenuation
curve. The distinction arises from the fact that very young stars
($t<10$ Myr), still embedded in their birth clouds, will suffer higher
attenuation than older populations which will only be attenuated by
the diffuse ISM \citep{fanelli88,keel93,calzetti94,cf00,wild11}. In
order to derive the intrinsic curve we assume that the attenuation
curves for young and old populations of a given galaxy are the same
($A_{\lambda, \rm {young}}/A_{V, \rm {young}}=A_{\lambda, \rm
  {old}}/A_{V, \rm {old}}$), but have different levels of
attenuation. Specifically, we use
$E(B-V)_{\rm old}=0.44E(B-V)_{\rm young}$, which is equivalent to
$A_{V, \rm {young}}/A_{V, \rm {old}}=2.27$, following the value for
the relationship between stellar and nebular reddening in
\citealt{calzetti00}. Assuming a factor of two greater (20 Myr) or
smaller (5 Myr) age demarcation between young and old populations does
not affect the results significantly. Note that even if both the young
and the old populations had the same intrinsic attenuation curves, as
assumed in this study, the resulting effective curve will be slightly
steeper than these intrinsic curves (by, on average,
$\propto \lambda^{-0.2}$), because more highly attenuated young
population dominates at shorter wavelengths, making the effective
curve steeper. We have verified that assuming, instead, a fixed law
for the young population and a variable slope for the older one does
not change, on average, the derived slopes of the dust attenuation
curve.

In this work we utilize the fine resolution model grid specifying SF
histories, metallicities and dust attenuations as in GSWLC-1, while
expanding the range and refining the resolution of the two parameters
specifying the dust attenuation curve. Specifically, we cover UV bump
amplitudes ($B$) from -2 to 6, in steps of 2 and power-law slope
deviations ($\delta$) with exponents from -1.2 to 0.4, in steps of
0.2. The range and steps were determined based on the range and
accuracy of these parameters in test and mock runs (Appendix
A.1). Unphysical negative value for the amplitude of the UV bump was
introduced to allow the mean of the relatively wide probability
distribution function to assume a value close to zero (see a similar
discussion in \citealt{salmon16}), and thus obtain an unbiased result
for galaxies that in reality lack the UV bump. Keeping the high grid
resolution as in GSWLC-1, while treating the dust attenuation curve
parameters as essentially free, results in 2.8 million distinct models
at each redshift. Models are calculated at 30 redshifts
($0.01<z<0.30$), separated by 0.01.

We refer to the physical parameters determined from this new, SED+LIR
fitting as GSWLC-2 (the medium-deep catalog would be referred to as
GSWLC-M2), and will make it publicly available from the project web
site\footnote{\url{http://pages.iu.edu/~salims/gswlc}}.

\section{Results} \label{sec:results}

In this section we first explore the demographics of dust attenuation
curves (``curves'') of different populations of galaxies, and then
present average curves in relation to curves from prior work,
together with analytical expressions for their construction.

We characterize the curves by parameters $\delta$ and $B$,
representing the power-law slope deviation with respect to the
Calzetti curve, and the strength of the UV bump in the total ($k$)
formulation of attenuation curve. Values $\delta<0$ signify curves
that are steeper than the Calzetti curve in the UV and blue
optical. For reference, the Milky Way bump in the extinction curve has
a mean strength of $B=3$. However, the contribution of the bump with
amplitude $B$ to attenuation at 2175 \AA\ will be smaller ($\sim3/4$) for
the modified Calzetti curve than for the MW extinction curve.

For the sample of star-forming galaxies, which will be our main focus,
the mean error of the dust attenuation curve slope deviation,
calculated as the standard deviation of the probability distribution
function, is 0.25, with 90 percentile range of errors extending from
0.1 to 0.4. The mean error of the determination of the bump amplitude
is 1.8, with 90 percentile range between 0.5 to 2.7. Mock SED fitting
(Appendix A.1) shows that despite the dispersion, both the slope and
the bump strength are recovered without significant systematics. In
what follows we will focus on average curve properties of a binned
distribution of galaxies, which are robustly measured due to the large
number of galaxies in the sample.

\subsection{Trends between dust attenuation curve parameters and
  galaxy physical parameters}

In this section we present average attenuation curve parameters for
galaxies of different stellar masses, as a function of various
physical and geometric parameters.

Figure \ref{fig:ssfrm_map} shows contour maps of average dust
attenuation curve parameters on sSFR--$M_*$ diagrams. Diagrams of
sSFRs vs.\ the stellar mass have emerged as an essential tool for
characterization of galaxy populations and their evolution
\citep{guzman97,pg03,bauer05,s07}. Diagrams featuring SFR normalized
by stellar mass (sSFR) are easier to interpret than the equivalent
SFR--$M_*$ diagrams, and are more fundamental than the related
color-magnitude diagrams. The upper panels of Figure
\ref{fig:ssfrm_map} show the full sample, consisting of both actively
star-forming (typically log sSFR$>-11$), and quiescent galaxies, while
the lower panels show only BPT-classified star-forming galaxies. The
white line represents the median sSFR trend of star-forming galaxies
(the star-forming sequence), and is repeated in the upper panels.

Focusing on the left two panels of Figure \ref{fig:ssfrm_map}, we see
that the average slopes of attenuation curves span a significant range
of values across the sSFR--$M_*$ plane. Attenuation curves are on
average the shallowest (lightest shade) for the most massive
star-forming galaxies, having slopes around that of the Calzetti curve
($\delta=0$). However, both the less massive star-forming galaxies as
well as the more quiescent galaxies have significantly steeper curves,
up to ($\delta\sim-0.9$), which is steeper than the slope of SMC
extinction curve (roughly $\delta=-0.45$). There are no regions with
average slopes shallower than the Calzetti curve. 

For star-forming galaxies (lower left panel) the primary trend in the
slope is the one with respect to the mass (higher mass = shallower
slope), but there is also a secondary trend with sSFR: at any given
mass, the curves tend to be steeper on both sides of the main sequence
(the shallowest slopes at a given mass are slightly below the main
sequence). In particular, a galaxy with high (s)SFR for its mass (a
``starburst'') will have a somewhat steeper curve than the galaxy of
the same mass and with more typical (s)SFR (lying closer to the white
line). We will show shortly that the trend with mass is a consequence
of the more fundamental dependence on optical opacity, while the trend
with sSFR is independent from it. Star-forming galaxies, like the
general population, on average have steep slopes (Section
\ref{ssec:comp}).

\begin{figure}
\epsscale{1.1} \plotone{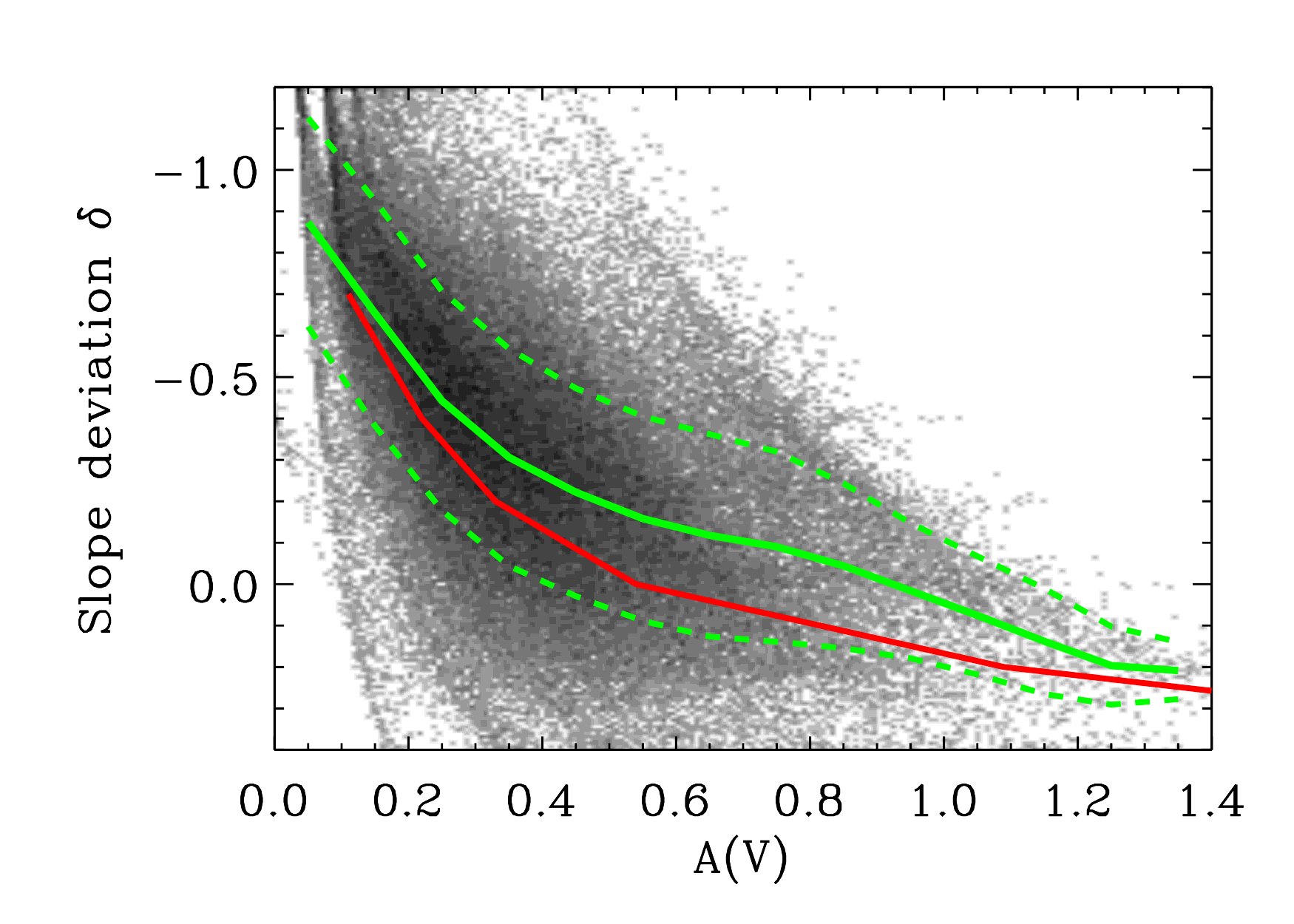}
\caption{Correlation between the dust attenuation slope and effective optical
  opacity. Attenuation curves of galaxies with higher optical opacity
  are systematically shallower. Slope is expressed as the exponent of
  the power-law deviation ($\delta$) with respect to the Calzetti
  curve ($\delta=0$). The average trend and the $1\sigma$ range around
  it are shown as green curves. The relation is in qualitative
  agreement with \citet{chevallard13} modelling results (red
  curve). Mild discretization of model grid parameters is
  visible. \label{fig:av_slope}}
\end{figure}

\begin{figure}
\epsscale{1.0} \plotone{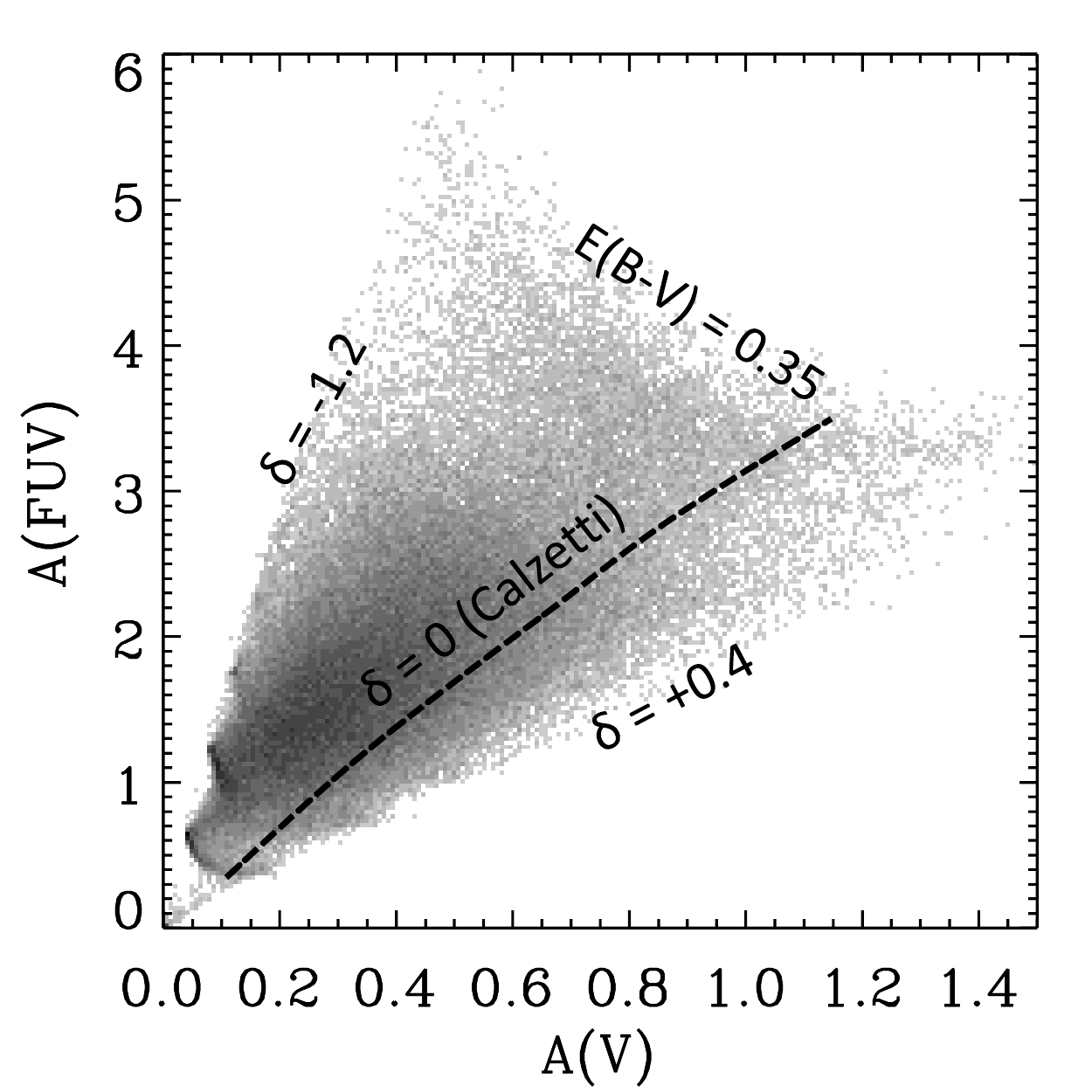}
\caption{Effective attenuation in FUV versus the attenuation in
  $V$. Because of the wide range of slopes of dust attenuation curves,
  the distribution is very wide. Upper and lower envelopes correspond
  to extrema in slopes. The locus of points for the Calzetti curve is
  indicated. Upper right envelope is defined by the galaxies with the
  highest reddening (labeled). Galaxies with similar $E(B-V)$run
  parallel to this label.\label{fig:afuv_av}}
\end{figure}

\begin{figure}
\epsscale{1.05} \plotone{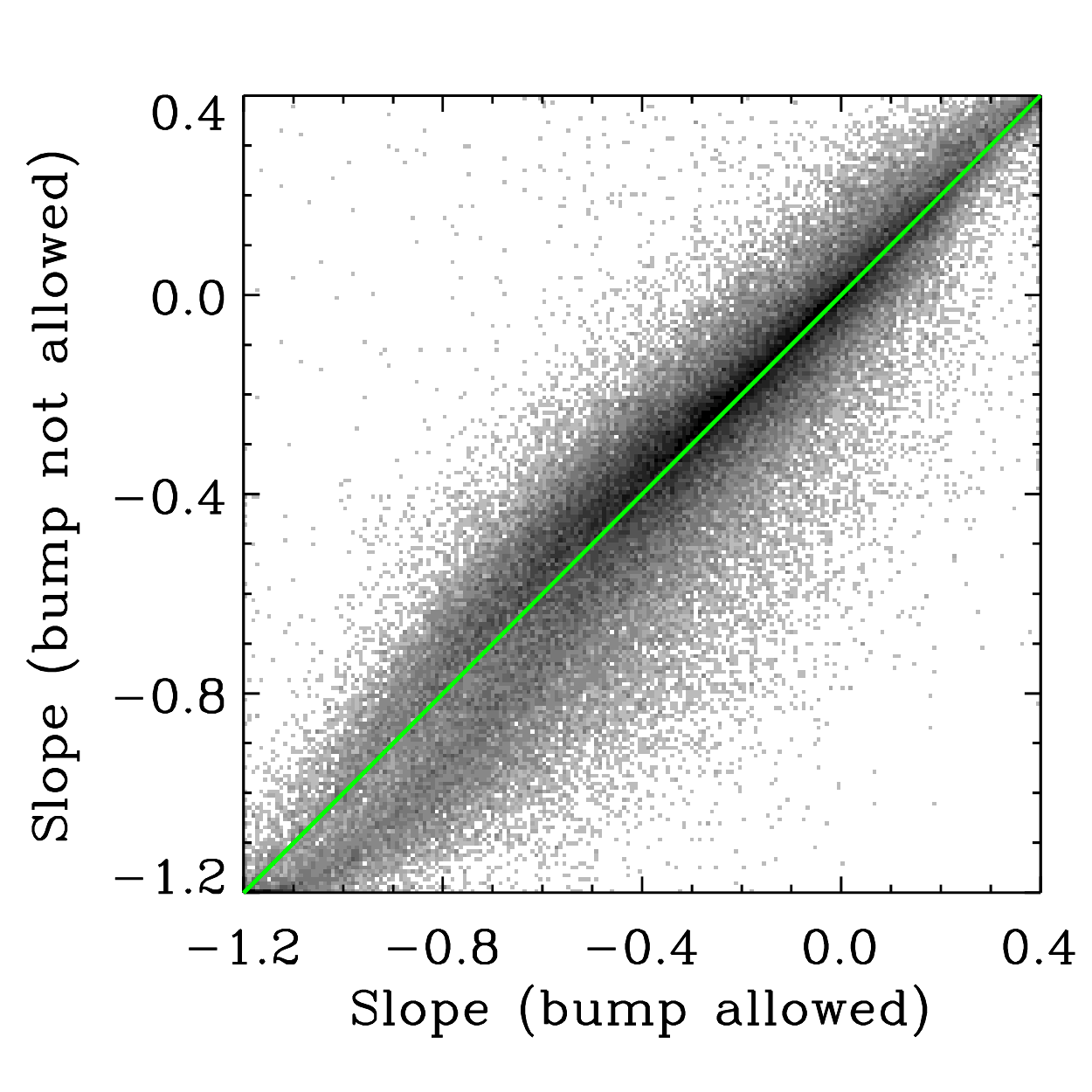}
\caption{Comparison of dust attenuation slopes (i.e., their deviations
  from the Calzetti curve) for the nominal SED-fitting run, and one in
  which the bump is not allowed. The slopes show no systematic offset
  (green line is 1:1 line), indicating that their steepness is not an
  artifact of allowing for a potentially poorly constrained
  bump. \label{fig:slope_comp}}
\end{figure}

\begin{figure*}
\epsscale{1.15} \plotone{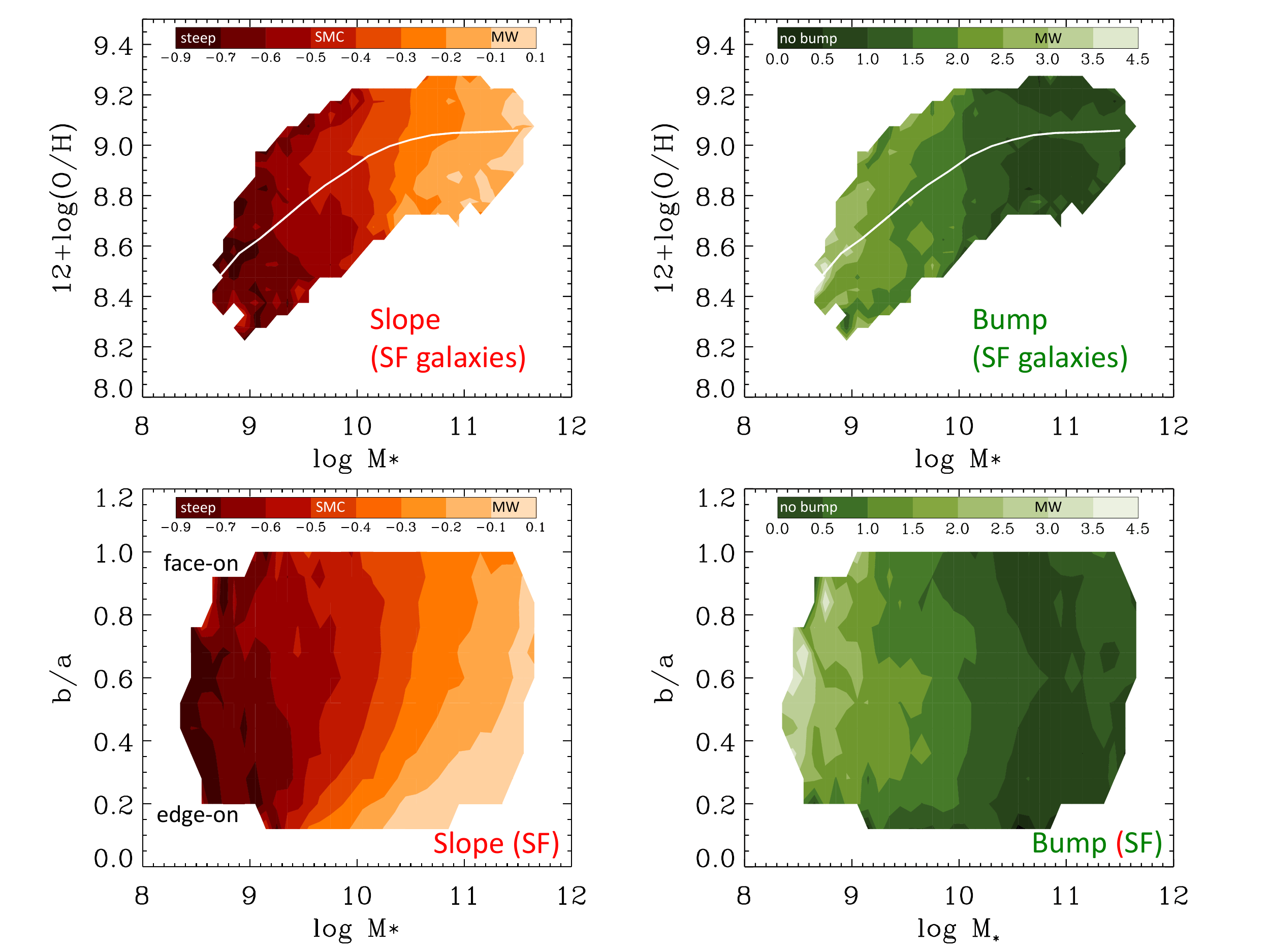}
\caption{Maps of dust attenuation curve parameters as a function of
  gas-phase metallicity (upper panels) and galaxy inclination (lower
  panels). The white lines in the upper panels are median trends,
  representing the mass-metallicity relation. There is no strong
  dependence of either the slope or the UV bump strength on
  metallicity, while the slope is shallower for more inclined massive
  galaxies. \label{fig:metinc_map}}
\end{figure*}

We now focus on the right panels of Figure \ref{fig:ssfrm_map}, showing
the amplitude of the UV bump in bins of sSFR--$M_*$ parameter
space. Again there is a range of values, from zero to MW-like. The bump is
most pronounced in star-forming galaxies of lower mass and for
quiescent galaxies. Massive star-forming galaxies ($\log M_*\sim11$)
are consistent with little or no bump. This suggests either that the MW is
atypical for galaxy of its mass, or that its extinction curve does not
reflect its attenuation curve if it were to be observed as an external
galaxy. The LMC, which also has a MW-like bump agrees better with
other galaxies of that mass. No group of galaxies have bumps
significantly stronger than the one in MW extinction curve. As in the
case of the slope, the strength of the bump exhibits a correlation
with sSFR, increasing on both sides of the main sequence. Unlike the
slope, the trends of bump strength against the mass or sSFR are not
the result of some more fundamental trend with the level of
attenuation, either in $V$ or in FUV. We conclude that, on average,
the UV bump in galaxies is moderate compared to the MW bump.

In the rest of the paper we focus on galaxies classified as
star-forming, for which the dust attenuation parameters are derived
with greater precision and which are observed at a wide range of
redshifts.

Next we explore, again in the form of contour maps, the dependence of
curve parameters on the level of attenuation in the optical ($V$) and
the far-UV (FUV). These attenuations are effective, arising from both
the young and old populations. Left panels of Figure \ref{fig:atten_map} show
slope deviations as a function of $A_V$ and $M_*$ in the upper panel, and
$A_{\rm FUV}$ and $M_*$ in the lower panel. The white line shows the
median attenuation for galaxies of different mass. While galaxies tend
to have higher attenuation as the mass increases, there is a
substantial scatter, especially for galaxies of higer mass. The slope
of the attenuation curve is strongly correlated with optical
attenuation, with galaxies with low $A_V$ values having steep curves
and the ones with high $A_V$ being shallow. Since the more massive
star-forming galaxies have higher optical opacities (the rising white
line in the upper left panel of Figure \ref{fig:atten_map}), the trend
between the slope and a mass seen in Figure \ref{fig:ssfrm_map} is
simply the result of the underlying trend between the slope and
$A_V$. At fixed $A_V$ there is no trend of slope versus mass (contours
are horizontal in Figure \ref{fig:atten_map} (upper left).

The trend between slope deviation and $A_V$ is shown separately in
Figure \ref{fig:av_slope}. The correlation is strong, and the scatter
at a given $A_V$ is comparable to slope determination error. In other
words, there may be little intrinsic scatter. From this we see that
only the galaxies with substantial optical attenuation ($A_V>0.8$)
tend to have, on average, curves as shallow as the Calzetti curve. The
dependence of the slope on the far-UV attenuation is to some degree
reversed: the slopes are steeper as the attenuation increases, or as
the mass decreases. The opposite sense of the trends may appear
paradoxical: one expects dust attenuation in the UV to follow the
attenuation in the optical. This would have been the case if galaxies
obeyed a universal attenuation law, but is not necessarily true when
there is a wide range of dust attenuation slopes, as is the case
here. This can be seen in Figure \ref{fig:afuv_av}, where
$A_{\rm FUV}$ is plotted against $A_V$. The upper envelope of points
corresponds to the steepest slopes, whereas the lower envelope
corresponds to shallowest slopes.

The UV bump, as mentioned, tends to be more prominent in galaxies with
steeper slopes, which, on average means lower $A_V$ and higher
$A_{\rm FUV}$. This agrees with the trends in the right panels of
Figure \ref{fig:atten_map}, and has been detected even at higher
redshifts \citep{kriek13,tress18}. An apparent correlation between the
bump intensity and slopes may raise a concern that the steep curves
that we are finding are an artifact of some underlying degeneracy with
the bump. To test for this possibility, we produce a run in which the
bump is not allowed, whereas all the other parameters of the SED
fitting are identical. If the slope and the bump are artificially
correlated, not allowing for the bump would force the slope to be
systematically different from the nominal slope. Figure
\ref{fig:slope_comp} compares the slopes from the two runs. There is
no systematic difference: the slopes remain on average the same
regardless of whether the bump was allowed or not. We conclude that
the bump is correlated with, but is not confounded with the slope.

A related concern is that non-zero bump amplitudes are just fitting
artifacts. We have performed a test to check if this may be the
case. We perform the fitting without the NUV band, using the models
without the UV bump. We, then look at the difference in the NUV
magnitude predicted by no-bump SED fitting and the actual NUV
magnitude. We find that the predicted no-bump magnitude is on average
brighter than the actual magnitude for galaxies with higher sSFRs, in
accordance with the bump amplitude derived from the SED
fitting. However, it should be pointed out that the effect on NUV
magnitude, even of a MW-like bump, is small (0.2 mag), and in the case
of our average bumps it is less than 0.1 mag.

In Figure \ref{fig:metinc_map} we look at the trends of attenuation
curve parameter versus gas-phase metallicity (upper panels) and
inclination (lower panels). The metallicity has been calculated using
the N2O2 method (the ratio of [NII]6584 to [OII]3727) and calibration
from \citet{kewley02}. This is the method of choice because it is less
sensitive to the ionization parameter than most other commonly used
methods \citep{kewley02}. For both the slope and the bump strength,
the contours are mostly vertical, meaning that there is not much
dependence on the metallicity. 

The trends involving galaxy inclination (derived as the simple axis
ratio from the exponential fit to galaxy's 2D profile; {\tt AB\_exp})
are shown in lower panels of Figure \ref{fig:metinc_map}. The bump is
largely insensitive to the galaxy orientation. There is some trend for
the slope of non-dwarf (presumably more disky) star-forming galaxies,
in the sense that face on galaxies have slightly steeper slopes. This
trend is not independent on the trend on optical opacity, and largely
goes away when $A_V$ is controlled for.

\subsection{Average dust attenuation curve: comparison with literature
curves} \label{ssec:comp}

In this section we present average dust attenuation curves for all
star-forming galaxies, for star-forming galaxies
binned by stellar mass, and for high-redshift analogs (starbursts). For each galaxy in these
groups we reconstruct individual
$A_{\lambda}/A_V$ curves from $\delta$ and $B$ and then average the
curves in small wavelengths bins from 912
\AA to 2.2 $\mic$. We additionally characterize these curves by the overall
power-law exponent $n$ (following \citealt{cf00}) determined from a
fit (in log of $\lambda$) to the curve in the range 912 \AA to 2.2 $\mic$:

\begin{equation}
A_{\lambda}/A_V = (\lambda/0.55\mic)^{-n} \label{eqn:pl}
\end{equation}

\begin{figure}
\epsscale{1.2} \plotone{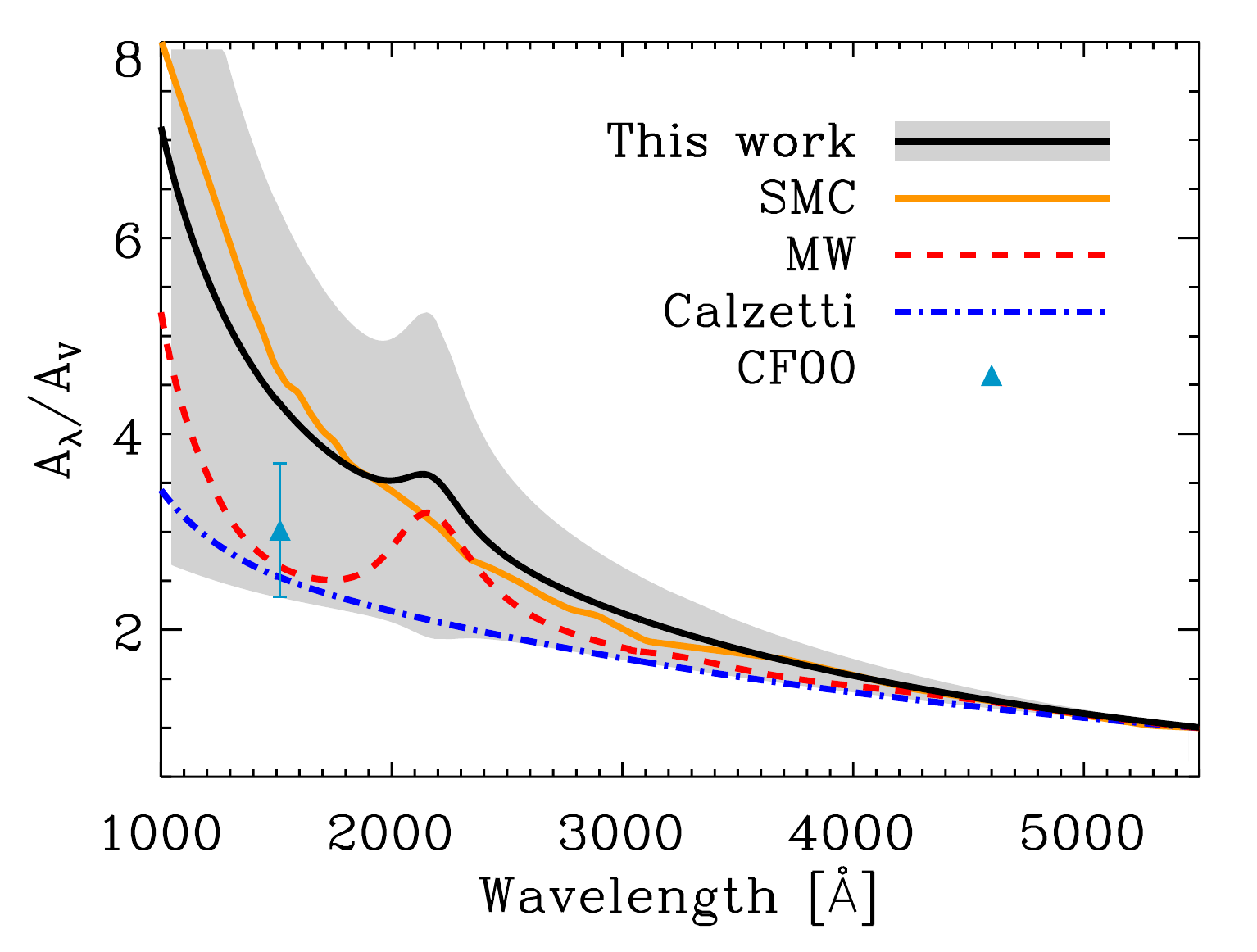}
\caption{Comparison of our average dust attenuation curve for
  star-forming galaxies (black line) to several canonical attenuation
  and extinction curves. Curves are normalized to $A_V$. Our average
  curve is steeper than the Milky Way \citep{cardelli89,odonnell94})
  (red dashed line) and \citet{calzetti00} curves (blue dot-dashed),
  and is more similar in slope to the SMC curve \citep{prevot84}
  (orange). The UV bump is 1/3 the Milky Way bump. Our curves of
  individual galaxies span quite a range---grey area shows $1\sigma$
  dispersion around the mean curve. The triangle shows the average and
  the $1\sigma$ range of curves obtained by the application of
  \citet{cf00} model with the intrinsic slope of
  $n=0.7$. \label{fig:dals}}
\end{figure}

Our average curve for star-forming galaxies is shown in Figure
\ref{fig:dals} as a black curve, with the gray region showing 1
$\sigma$ dispersion across the sample. Dispersion is zero at 5500 \AA\
by construction. We see that the average curve ($n=1.15$) is almost as
steep as the SMC curve (\citealt{prevot84}; yellow curve,
$n=1.2$). The average curve exhibits a moderate UV bump, whose
contribution to attenuation at 2175 \AA\ is roughly 1/3 that of the
contribution of the MW bump to its curve at 2175 \AA. The effect of
such bump on the observed NUV magnitudes is $\sim 0.07$ mag. The Milky
Way extinction curve from \citet{cardelli89} (red curve) is less steep
overall ($n=0.9$), but soars at the far-UV and has more pronounced UV
bump. Given that MW and SMC are extinction, not attenuation curves,
they are of limited value for drawing conclusions as to hoe well the
MW or the SMC follows our mean curve. The Calzetti curve is similar to
the MW curve except that it is shallower in the far-UV and does not
exhibit the bump, with an overall power-law slope of $n=0.75$. The
range of curves found in this work extends from Calzetti-like curves
at the shallow extreme, up to the steeper-than-SMC curves ($n=1.6$) on
the other end. We also show a range of $A_{\rm FUV}/A_V$ values
corresponding to SDSS/\galex\ galaxies fitted to a suite of model SEDs
based on \citet{cf00}. \citet{cf00} models, having different $A_V$
normalizations for young and old populations and the intrinsic slope
of $n=0.7$ naturally produce a range of effective attenuation curve
slopes, which is, however, still smaller and on average less steep
($n=0.9$) than what we find for the same galaxies (c.f.\
\citealt{hayward15}).

\begin{figure}
\epsscale{1.2} \plotone{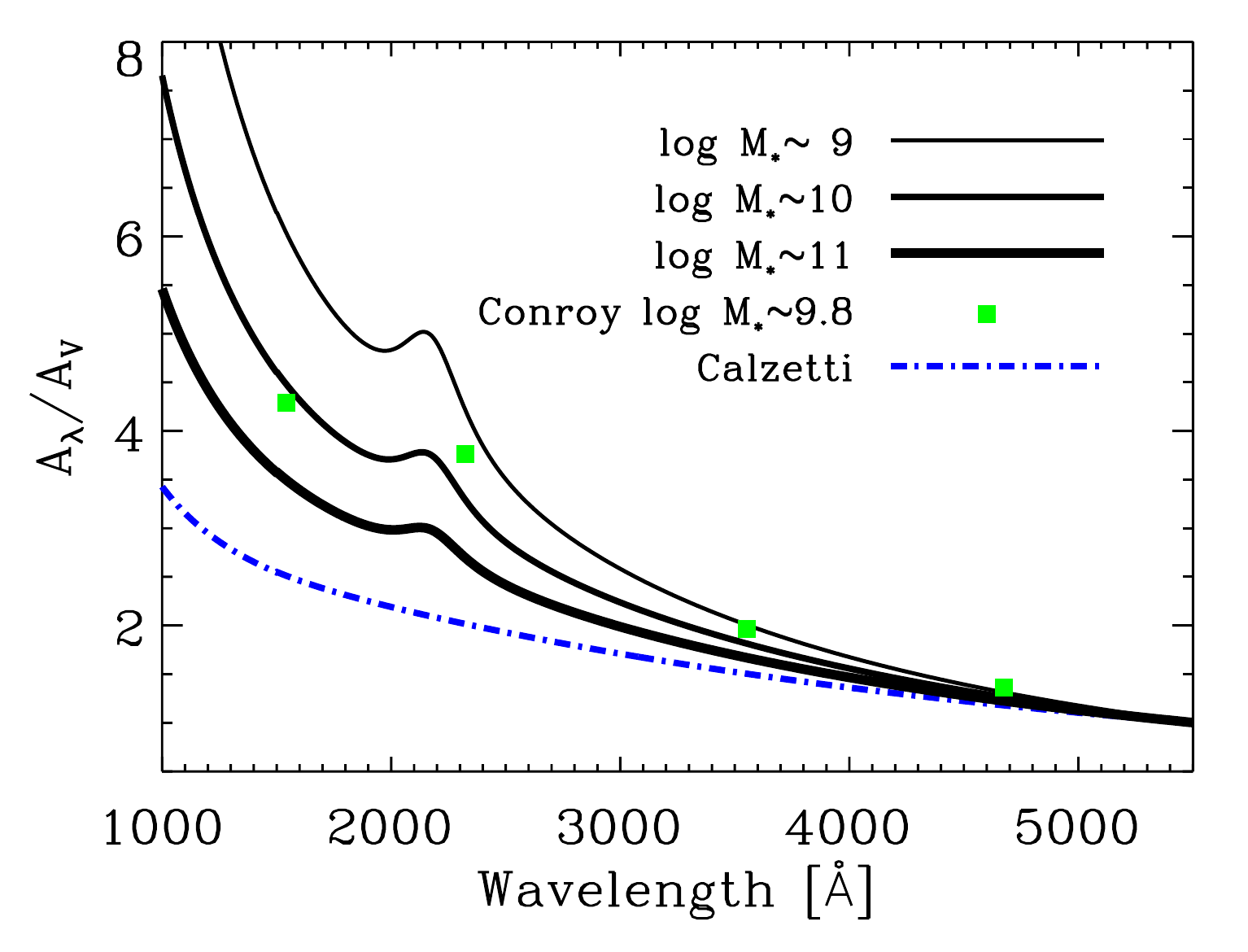}
\caption{Comparison of our average dust attenuation curves for galaxies of
  different mass (black lines) with several curves from the literature.
  Curves of lower-mass galaxies tend to be steeper, which we find to
  be correlated with their lower dust content. There is a good
  agreement with the points (green squares) corresponding to the
  attenuation curve of \citet{conroy10uv}, derived from UV-optical
  photometry of a sample of $9.5<\log M_*<10$. The relative
  contribution of the UV bump declines with mass.
  galaxies. \label{fig:dals_mass}}
\end{figure}

\begin{figure}
\epsscale{1.2} \plotone{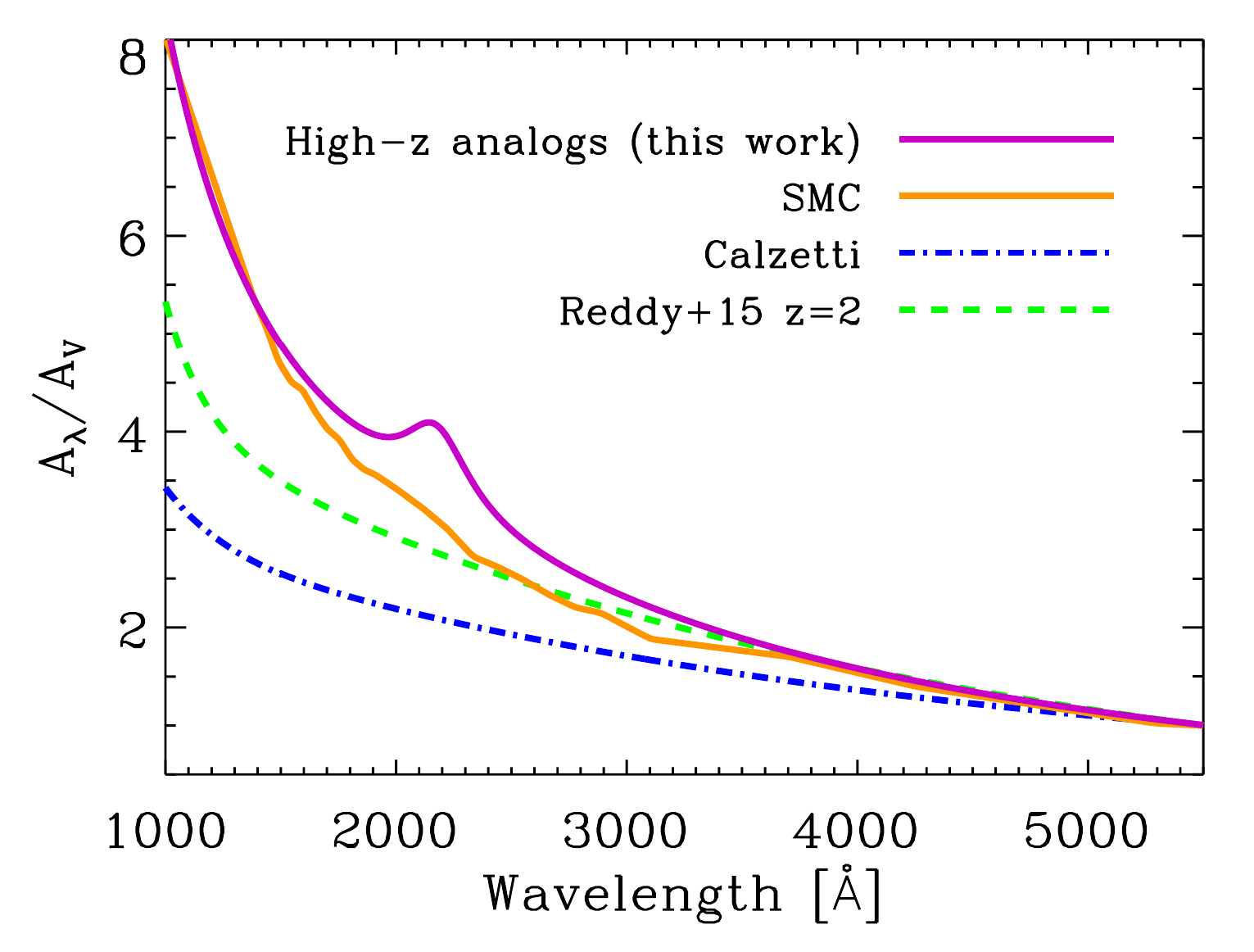}
\caption{Comparison of average dust attenuation curves for our sample
  of high-redshift analogs (galaxies lying $>0.5$ dex above the local
  main sequence; purple line) and several curves from the literature,
  including \citet{reddy15} (green dashed curve).  Local analogs
  (starbursts) have steep curves, similar to the SMC slope (orange,
  \citealt{prevot84}), but with more extinction in the near-UV, partly
  due to the moderate UV bump. From this we argue that high-redshift
  galaxies have a range of curves that are on average as steep or even
  steeper than the local galaxies and therefore as steep or steeper
  than the SMC curve. \label{fig:dals_dssfr}}
\end{figure}

In Figure \ref{fig:dals_mass} we show average curves for star-forming
galaxies, now split in three bins of stellar mass (black curves). The
curve for the massive galaxies ($\log M_*\sim11$) is the shallowest
($n=1.0$), but still not as shallow as the Calzetti curve. The average
curve for less massive galaxies is steeper still, with a moderate UV
bump.  \citet{conroy10uv} have used UV (\galex) and optical (SDSS)
colors to constrain the dust attenuation curve of a $9.5<\log M_*<10$
sample and have arrived at values shown in Figure \ref{fig:dals_mass}
as green squares. There is a good overall agreement with our results
for galaxies of comparable mass.  We can derive an expression between
the average slope deviation and the mass as:

\begin{equation}
\delta = -0.38+0.29 (\log M_*-10)
\end{equation}

\noindent from which the overall power-law slope can be obtained as $n =
-\delta+n_{\rm Cal}=-\delta+0.75$.

We also derive an average curve for high-redshift analogs. Following
\citet{s15}, we define high-redshift analogs as galaxies with large
positive offset from the main sequence, specifically
$\Delta \log {\rm SSFR} >0.5$ (the results are the same if a more
extreme criterion of at least 1 dex of offset is adopted.)  It is
worth pointing out that selecting high-redshift analogs (or, more
generally, starbursts) based on the offset form the main sequence is
in close accordance with how galaxies evolve (the shifting
normalization of the main sequence). Using instead some cutoff in sSFR
would preferably select low-mass galaxies (Figure
\ref{fig:ssfrm_map}), whereas a cut in SFR would preferably select
high-mass galaxies. The sample of \citet{calzetti94} lies on average 1
dex above the local main sequence, i.e., occupies a similar position
as our $\Delta \log {\rm SSFR} >0.5$ selection.

The resulting curve for high-redshift analogs is shown in Figure
\ref{fig:dals_mass} in purple. It is similar to the curve for the
general population of intermediate-mass galaxies, with SMC-like far-UV
rise ($n=1.25$), but somewhat higher attenuation in the near-UV
(partly due to a moderate bump). The curve of \citet{reddy15} derived
from $z\sim2$ galaxies using the methodology of
\citet{calzetti94,calzetti00} is less steep than ours (green dashed
curve). Their {\it selective} curve is essentially identical to the
Calzetti curve in the UV/optical region, but is effectively steeper in
absolute normalization because of the lower $R_V$ (2.505 vs.\
4.05). Our steep curve for high-$z$ analogs agrees better with a
recent re-assessment of $z\sim2$ observations by \citet{reddy17}, who
find that the SMC-like curve gives a better match to the observed
IRX-$\beta$ relation than shallower curves. What is significant is
that we find that such a steep curve is needed for local starbursts as
well, and that such galaxies have slightly steeper curve, on average,
than the general population of local star-forming galaxies. We discuss
this result further in Section \ref{sec:disc}. A correlation between
the slope deviation and the stellar mass is present for high-redshift
analogs as well, and is given by:

\begin{equation}
\delta = -0.45+0.19 (\log M_*-10)
\end{equation}


\begin{deluxetable*}{l r r r r r r r r} 
 \tablecaption{Functional fits of dust attenuation curves. \label{table:pms}}
\tablewidth{0pt}
\tablenum{1}
 \tablehead{
   \colhead{Sample}      &
   \colhead{$R_V$}      &
   \colhead{$B$} &
   \colhead{$a_0$} &
   \colhead{$a_1$} &
   \colhead{$a_2$} &
   \colhead{$a_3$} &
   \colhead{$\lambda_{\rm max}(k>0)$} &
   \colhead{$n$} \\
   }
\startdata
Star-forming galaxies & 3.15 & 1.57 & $-4.30$ & 2.71 & $-0.191$ & 0.0121 & 2.28
& 1.15\\
$\quad 8.5< \log M_* \leqslant 9.5$ & 2.61 & 2.62 & $-3.66$ & 2.13 & $-0.043$ & 0.0086 & 2.01
& 1.43\\
$\quad 9.5< \log M_* \leqslant 10.5$ & 2.99 & 1.73 & $-4.13$ & 2.56 & $-0.153$ & 0.0105 & 2.18
& 1.20\\
$\quad 10.5< \log M_* \leqslant 11.5$ & 3.47 & 1.09 & $-4.66$ & 3.03 & $-0.271$ & 0.0147 & 2.45
& 1.00\\
High-$z$ analogs & 2.88 & 2.27 & $-4.01$ & 2.46 & $-0.128$ & 0.0098 & 2.12
& 1.24\\
$\quad \log M_* \leqslant 10$ & 2.72 & 2.74 & $-3.80$ & 2.25 & $-0.073$ & 0.0092 & 2.05
& 1.38\\
$\quad \log M_* > 10$ & 2.93 & 2.11 & $-4.12$ & 2.56 & $-0.152$ & 0.0104 & 2.09
& 1.19\\
Quiescent galaxies & 2.61 & 2.21 & $-3.72$ & 2.20 & $-0.062$ & 0.0080 & 1.95 & 1.35\\
\enddata
\tablenotetext{}{$B$ is the amplitude of the UV bump, while $a$ are
  the coefficients of the polynomial fit to the total attenuation
  curve ($k$ formulation; Eqn.\ \ref{eqn:poly},
  \ref{eqn:drude}). Curves have positive values at
  $\lambda<\lambda_{\rm max}$, beyond which they should be
  annulled. Last column gives $n$, the exponent of the power-law
  approximation of the attenuation curve (Eq.\ \ref{eqn:pl}).
  Star-forming galaxies are selected using the BPT emission-line
  diagnostic diagram, which yields galaxies with log
  sSFR$>-11$. High-$z$ analogs are the subset of star-forming galaxies
  lying more than 0.5 dex above the median star-forming sequence
  ($\Delta \log {\rm SSFR} >0.5$). Quiescent galaxies are galaxies
  with log~sSFR$<-11$. Star-forming galaxies and high-$z$ analogs are
  additionally split by stellar mass.}
\end{deluxetable*}

\subsection{Functional fits}\label{ssec:func}

We fit our curves with polynomial functions following the formalism of
\citet{calzetti00}, where the fit is performed in total formulation of
the curve, from which an absolute curve can obtained by dividing by
$R_V$ (Section \ref{ssec:nom}). \citet{calzetti00} present their curve
as a piecewise fit: a third-order polynomial in $\lambda^{-1}$ for the
UV+blue optical region, and a linear function in $\lambda^{-1}$ for
the red optical+near-IR region, with the split between two regimes at
0.63 $\mic$. Considering that the level of uncertainty far exceeds the
refinement provided by splitting the curve in two parts, we derive
functional fits based on the entire $0.912<\lambda<2.2\,\mic$ range,
using the third-order polynomial:

 \begin{eqnarray}
k_{\lambda} & =&
 a_0+a_1\lambda^{-1}+a_2\lambda^{-2}+a_3\lambda^{-3}+D_{\lambda}(B)+R_V, 
\label{eqn:poly}
\end{eqnarray}

\noindent to which we add the UV bump as a Drude profile
$D_{\lambda}(B)$, with amplitude $B$ and fixed central wavelength (2175 \AA)
and width (350 \AA):

 \begin{eqnarray}
D_{\lambda}(B) & = &
                  \frac{B\lambda^2(0.35\mic)^2}{[\lambda^2-(0.2175\mic)^2]^2+\lambda^2(0.35\mic)^2}
\label{eqn:drude}
\end{eqnarray}

\noindent For the average curve of all star-forming galaxies we get
$R_v=3.15$, and the fit: 

\begin{eqnarray}
k_{\lambda} & =&
 -4.30+2.71\lambda^{-1}-0.191\lambda^{-2}+0.0121\lambda^{-3}+D_{\lambda}(1.57)+3.15, \nonumber \\&&0.09\leqslant
 \lambda <2.2\, \mic 
\end{eqnarray}

\noindent Note that the fit produces negative values for
$\lambda>2.28\,\mic$, so it is recommended to replace it by zero
values in that regime.

Alternatively to using the polynomial fits, one can use the expression
for the modified Calzetti law (Eqn.\ \ref{eqn:mod1}, \ref{eqn:mod2})
with the values $\delta=-0.4$ and $B=1.3$. A somewhat cruder but still
useful characterization of the curve is a simple power law form (Eq.\
\ref{eqn:pl}).  Considering that the UV bump is relatively modest in
average curve, neglecting it should not lead to large systematics.
Thus, our mean law can be approximated with an absolute curve with the
best-fitting power-law exponent of $n=1.15$.

Considering the diversity of dust attenuation curves, it is useful to
have functional forms for specific subsamples. Table \ref{table:pms}
lists polynomial coefficients for the fits of star-forming galaxies
split by stellar mass. Furthermore, we present fit parameters for
high-redshift analogs: all, and split by mass, and for the quiescent
galaxies ($\log {\rm sSFR}<-11$, any BPT class).

It should be pointed out that unless the IR dust emission constraints are
used, one should refrain from leaving the curve as a free parameter
in the SED fitting, as the resulting ill-constrained values may bias
the derived values of SFRs. In such cases it is preferable to use a
fixed, but appropriate, average curve.

\section{Discussion} \label{sec:disc}

The main results of our work are, first, that galaxies exhibit a very
wide range of attenuation curves, and second, that the slopes are on
average steep, both for normal star-forming galaxies, and, to an even
greater degree, for galaxies above the main sequence, i.e., starbursts
or high-$z$ analogs. We have performed a number of tests (effective
vs.\ intrinsic curve, different old/young $A_V$ normalizations,
different old/young transition timescales, different dust absorption
factors for ionizing photons, fitting with and without the bump,
different assumed input errors, Chabrier vs.\ Salpeter IMF,
(non)correction of emission lines, \citet{m2005} vs.\ \citet{bc03} SPS
models, different SF histories, fitting with and without the FUV) and
the need for overall steep curves with moderate bumps remains.

Our results agree with other studies that employ model-based methods,
which point to generally steep curves with moderate bumps (less than
$B_{\rm MW}=3$) at both low and high redshifts. We review these
results, first focusing on low redshift.  \citet{conroy10uv}, using
UV/optical photometry of samples of galaxies at different
inclinations, find a steep slope with a bump of $B=2.4$, consistent
with what we find for galaxies of similar mass (Figure
\ref{fig:dals_mass}). \citet{burgarella05}, using UV to far-IR SED
fitting, also find on average a steep ($n=1.1$) and moderately bumpy
slope for a sample of $\sim100$ local galaxies. Most recently,
\citet{leja17} used UV-to-mid-IR SED fitting of a sample of 129 nearby
galaxies to constrain physical parameters including the slope of the
attenuation curve. They use similar methodology to ours, except that
the determination of $L_{\rm IR}$ is not decoupled from overall SED
fitting. They find the same range of slopes as we do
($-1.2<\delta<0.4$), but on average somewhat shallower, consistent
with the fact that their sample contains, on average, more massive
galaxies. We find that the dust attenuation curves tend to be steep
also below the main sequence, among the quiescent galaxies. Recently,
\citet{viaene17} obtained an optical attenuation curve of a nearby
lenticular, star-forming galaxy with prominent dust lanes, one of very
few individual galaxies with determined attenuation curve.  They find
a steep curve with a slope deviation of $\delta =-0.43$, in agreement
with our results for galaxies lying below the main sequence (Figure
\ref{fig:ssfrm_map}, upper left).

Our results also agree with the results of studies that
apply model-based methods to a general population of higher-redshift
galaxies. \citet{kriek13}, using stacked SEDs of medium and broad
band photometry of galaxies at $0.5<z<2$, find an average slope of
$\delta=-0.2$ and an average bump of $B=1$, with both parameters
spanning a range of values (their Fig.\ 2). Their sample is restricted
to galaxies with moderate to high optical attenuations ($A_V>0.5$),
which have shallower slopes and weaker bumps (Figure
\ref{fig:atten_map}, \ref{fig:av_slope}). When we isolate
$A_V>0.5$ galaxies, the average values of curve parameters agree
with those found by \citet{kriek13}. Evidence for a moderate bump
($0.5<B<2.5$) and a range of steep curves ($-0.7<\delta<0.1$) has also
been found in the SED analysis of a $1<z<2.2$ sample by
\citet{buat12}, as shown in their Fig.\ 8. A similar range of slopes,
though on average somewhat shallower than ours, was obtained in the
SED analysis of rest-frame UV to optical data of $z\sim2$ galaxies
\citep{salmon16}. 

A strong correlation between attenuation curve slope and optical depth
(Figure \ref{fig:av_slope}) is a result with important implications,
and it also impacts the underlying assumption of the comparison
method, which we will discuss shortly. Our results match the
predictions of \citet{chevallard13}, based on radiative transfer
models combined with realistic dust geometries and the two-component
(birth clouds/diffuse ISM) model \citep{cf00}. In particular the $A_V$
point at which the dependence of slopes on $A_V$ changes its character
from rapid (at $A_V<0.4$) to less rapid at ($A_V>0.4$) matches the
break in our trend (Figure \ref{fig:av_slope}. The existence of a
dependence of the attenuation curve slopes on $A_V$ can also be
inferred from the SED fitting results at higher redshift (Fig 12.\ of
\citealt{arnouts13}), and has been reported recently in \citet{leja17}
for a sample of 129 nearby galaxies.

\citet{chevallard13} provide the following physical picture for this
correlation. The steepness of an attenuation curve at small optical
depths is the result of the dominance of scattering over absorption,
coupled with the fact that scattering is more forward directed at
shorter wavelengths whereas it is more isotropic at longer
wavelengths. As the optical depth increases, absorption becomes more
dominant than scattering, and the curve becomes shallower 
(grayer). 

We show that the dependence of the slope of the curve on $A_V$ leads
to an apparent dependence on the stellar mass. A similar result, that
$\log M_*<10$ galaxies have steeper curve than more massive galaxies,
was recently obtained in the analysis of a large sample of $z\sim2$
galaxies by \citet{reddy17}.

\citet{chevallard13} furthermore show that the dependence of the slope
on $A_V$ is the same irrespective of whether the $A_V$ is driven by
different levels of intrinsic (face-on) attenuation, or is the result
of inclined viewing geometry. Regarding the latter, they predict that
attenuation slopes should steepen at $b/a>0.4$ (going from
edge-on to face-on), in agreement with our results for disk-geometry
galaxies (Figure \ref{fig:metinc_map}, lower left panel).

The strength of the UV bump also shows some correlation with $A_V$
(Figure \ref{fig:atten_map}, upper right panel). Similar behavior has
been recently seen by \citet{hagen17}, in the analysis of SMC dust
curves based on pixel SED fitting of UV (from {\it Swift}), optical
and 3.6 and 8 $\mic$ photometry. Interestingly, this study finds that
SMC exhibits a mild UV bump, on average 1/5 as strong as the MW
extinction curve bump. Furthermore, they show that the bump amplitude
is highest ($B\sim1.5$) when $A_V\sim0$, dropping rapidly to $B=0.3$
at $A_V=0.4$, in agreement with our results.

Furthermore, we find that neither the slope of the attenuation curve
nor the UV bump depend on gas-phase metallicity (Figure
\ref{fig:metinc_map}, upper panels). For the slopes this confirms,
using much larger sample, the results of \citet{calzetti94}, where no
systematic difference was found between the lower and higher
metallicity galaxies in plots of UV power-law index vs.\ the Balmer
optical depth. This lack of dependence for both the slope and the bump
places interesting constraints on the modeling of dust properties and
for understanding the nature of the dust component responsible for the
bump (e.g., \citealt{mathis94}). 

Overall, our results challenge the view, based on MW, LMC and SMC
extinction curves, that less massive galaxies have a weaker bump,
supposedly due to a lower metallicity. The results (Figure
\ref{fig:ssfrm_map}) also prompt us to reconsider the idea that
galaxies with high sSFRs have UV radiation field that lead to the
destruction of the carriers of the UV bump
\citep{fischera11}. According to this scenario the UV bump may be
prominent in normal SF galaxies but not in starbursts. The absence of
the bump in starbursts is supported by a non-detection of a strong (MW
or LMC-like) bump in \citet{calzetti94} analysis of {\it IUE} UV
spectra. However, it is notable that the UV bump is apparently absent
also in the {\it IUE} UV spectra of more normal SF galaxies from
\citet{kinney93} parent sample. Out of 140 galaxies in Kinney atlas
only one shows somewhat prominent UV bump feature (NGC 7714), a galaxy
that happens to be a starburst. The {\it IUE} UV spectra tend to be
noisy in the near-UV, while the expected effect from the bumps we find
is quite moderate (1/3 of MW), which may explain the difficulty of the
detection of the bump in local galaxies. On the other hand, the
stacking of high S/N observed-frame optical spectra by
\citet{noll07,noll09}, for their sample of exquisitely bursty galaxies
(log sSFR$\sim-8$), reveals strong evidence for a moderate UV bump in
1/3 of their $1<z<2.5$ sample, in agreement with our results.

While a growing number of studies show that average attenuation curves
are steep and moderately bumpy at lower and higher redshifts, the
question remains as to why we find that local starbursts
(high-redshift analogs) have even steeper curves than the ``normal''
star-forming galaxies, in clear contrast to the shallow Calzetti
curve.  We propose that there is a systematic difference in the slopes
derived by ``model-based'' and by ``comparison'' methods, especially
if the latter are based on the use of Balmer decrement as a proxy for
stellar attenuation.

\begin{figure}
\epsscale{1.2} \plotone{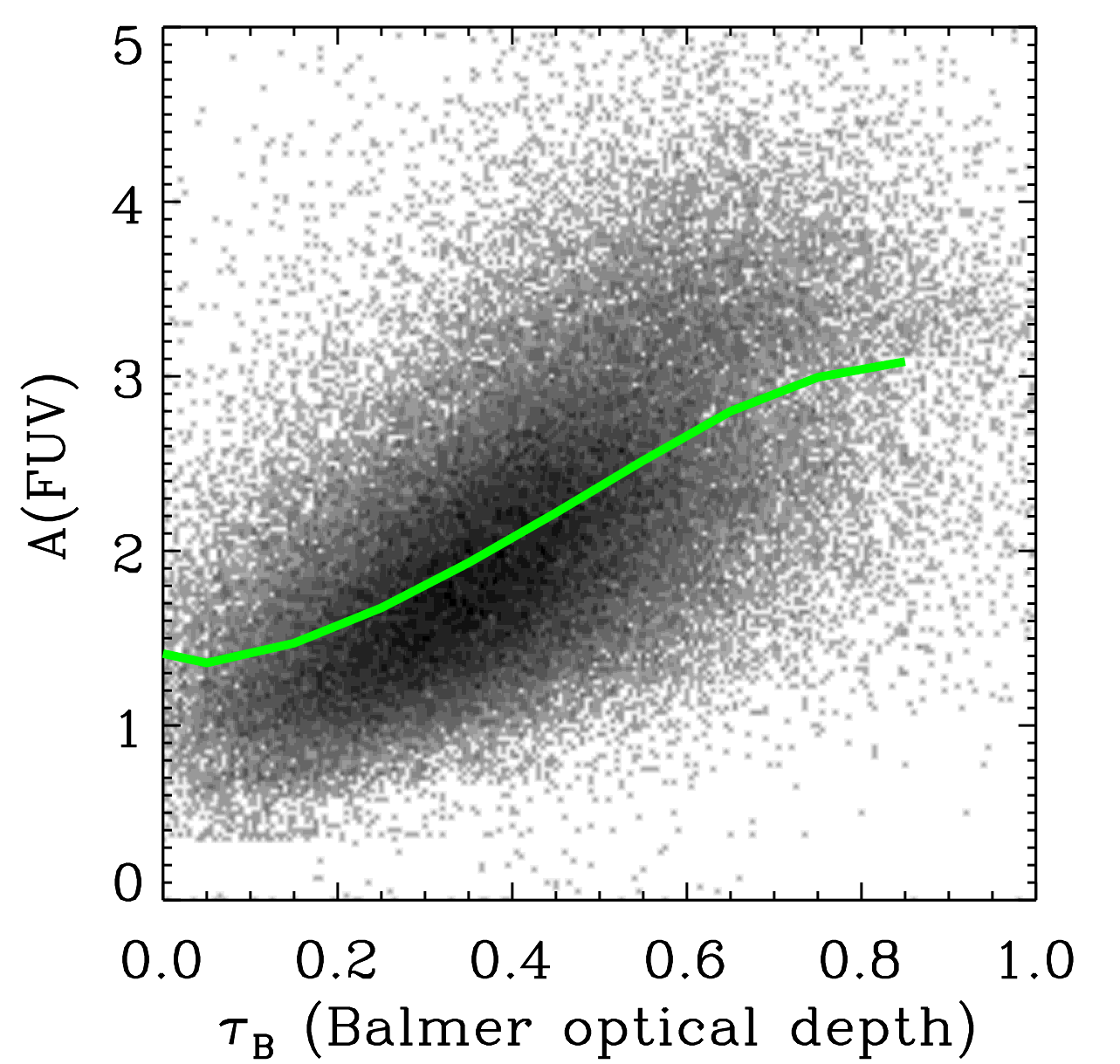}
\caption{Comparison of attenuation affecting the far-UV continuum
  light and the Balmer optical depth affecting the HII regions. While
  the two attenuations are correlated, there is a large degree of
  intrinsic scatter between the two, such that the identification of
  unreddened galaxies using the Balmer decrement results in
  galaxies that on average have a substantial dust attenuation (1.3
  mag in FUV). FUV attenuation is derived from our SED fitting and has a
  typical error of 0.3 mag (i.e., it does not dominate the scatter),
  whereas the Balmer decrement optical depths are based on SDSS
  spectra. Green curve represents the trend of binned
  averages. \label{fig:afuv_tau}}
\end{figure}

First, in order to check whether our steep slope of high-redshift analogs is
due to the differences in samples, we perform IR-luminosity
constrained SED fitting on \citet{calzetti00} galaxies that lie in
SDSS DR10 footprint, and have requisite photometry from \galex\ and
\wise. We take UV and optical photometry, as well as redshifts, from
NASA-Sloan Atlas (v0\_1\_2; \citealt{blanton11}), which, by collecting
redshifts from a variety of sources (including IPAC NED) has a more
complete redshift coverage at $z<0.05$ than the SDSS spectroscopic
catalog alone. Furthermore, NASA-Sloan-Atlas photometry of SDSS images
is optimized for extended objects. The resulting sample contains 14
out of 33 galaxies used in \citet{calzetti00}. On average, these
galaxies lie 1 dex above the main sequence, similar to our high-$z$
analogs. The resulting curves of these 14 galaxies are steep
($\delta=-0.56\pm0.4$, or $n=1.3\pm0.4$ for the average and the
1-$\sigma$ scatter), steeper on average than the SMC curve, and even
slightly steeper than the curve of our high-redshift analogs shown in
Figure \ref{fig:dals_dssfr}. The average of the total to selective
extinction in $V$ is $R_V=2.74$. The UV bump that we find for this
sample is very modest (somewhat less pronounced than for our high-$z$
analogs), in qualitative agreement with \citet{calzetti94}.

The reasons for the large discrepancy in slopes with respect to
\citet{calzetti00} for the same sample are probably methodological
rather than stemming from the data used (e.g., integrated photometry
vs.\ spectroscopy). As mentioned in Section \ref{sec:intro}, the
method of \citet{calzetti94} consists of a comparison of UV/optical
spectra of galaxies having different levels of attenuation, as
determined from the {\it nebular} emission (the Balmer decrement).
Once normalized to $E(B-V)=1$, this method produces the {\it
  selective} curve ($k_{\lambda}-R_V$; Section \ref{ssec:nom}). The
{\it total} curve requires the determination of $R_V$, through the use
of IR data (i.e., the energy balance), or by referring to nearly
unattenuated near-IR photometry. The total curve for
\citet{calzetti94} sample has been established subsequently, in
\citet{calzetti00}, by keeping the selective curve the same as in the
original work and solving for $R_V$. 
 
We point out two issues of relevance to the comparison method of
\citet{calzetti94} that may be the source of this discrepancy, of
which we believe the second is more important. The comparison method
looks at the SEDs of reddened galaxies with respect to the SEDs of
galaxies that are considered to be dust free. Galaxies are selected as
unreddened usually based on the nebular Balmer decrement, as a proxy
for continuum attenuation. First, as pointed out by \citet{wild11} and
\citet{chevallard13}, the underlying assumption of the comparison
method is that the attenuation curves do not vary with the dust
content, as characterized, for example, by optical opacity ($\tau_V$
or $A_V$). The validity of this assumption has not been fully
established in previous studies. Our results (Figure
\ref{fig:av_slope}) and theoretical analysis of \citet{chevallard13}
suggests that this requirement is not fulfilled. Because in the
comparison method the high-opacity (dusty) galaxies have the greater
leverage in the derivation of the curve (with weight being
proportional to $\Delta \tau_V$; \citealt{wild11}), the resulting
attenuation curve will be biased towards the shallower slopes of
high-opacity galaxies.

The second, and likely more significant factor is that the comparison
method relies on the use of {\it nebular} attenuation (the Balmer
decrement) to place galaxies in different attenuation
categories. However, the nebular extinction is a poor proxy for
continuum attenuation, due to a high intrinsic dispersion between
these quantities, as demonstrated in Figure \ref{fig:afuv_tau}. This
potential issue with the comparison method was pointed out by
\citet{cf00}. Based on SDSS emission line data we find that the
galaxies with $\tau_{\rm Balmer}<0.1$ (that would be considered
unattenuated), actually have significant average continuum attenuation
of $A_{\rm FUV}=1.3$ (Figure \ref{fig:afuv_tau}).  Likewise, the
galaxies considered to be highly attenuated according to Balmer
decrement ($\tau_{\rm Balmer}\sim0.7$) have an average
$A_{\rm FUV}=3.0$, well below the actual most opaque galaxies
($A_{\rm FUV}\sim4.5$). Thus the difference in attenuations between
galaxies selected as dusty and the ones assumed to be dust-free is
smaller by more than a factor of two than what is assumed based on the
Balmer decrement. Deriving the attenuation curve by dividing the SED
of a redenedded galaxy by an SED of a galaxy that is not truly
unreddened will result in a shallower slope.

More detailed investigation of systematics resulting from methodology
will be important for future work. Interestingly, other studies that
use the comparison method based on Balmer decrements also obtain
shallow slopes of dust curves (either selective or absolute),
regardless of whether the population consists of starbursting or more
normal galaxies. Our results do agree on the point that there is no
large difference in average curves of normal and starbursting (and
likely high-redshift) galaxies, but here we find that those curves
are, on average, systematically steeper, consistent with other studies
that are based on model-based approaches.

\section{Summary} \label{sec:conc}

The paper presents constraints on dust attenuation
curves for a very large sample of galaxies in the local universe,
allowing for detailed statistical characterization. Our main findings are as
follows:

\begin{enumerate}

\item Galaxies in the local universe exhibit a wide range of dust
  attenuation curve slopes (power-law exponent range $\sim1$), from
  shallow slopes similar to the slope of the Calzetti curve, to slopes
  significantly steeper than the SMC extinction curve.
\item On average, local star-forming galaxies have steep curves,
  almost as steep as the SMC extinction curve.
\item The steepness of the curves is most strongly correlated with the
  optical opacity ($A_V$), with higher
  opacity implying a shallower curve, confirming \citet{leja17}, and in
  agreement with the predictions of radiation transfer models with
  realistic geometry \citep{chevallard13}.
\item Slopes tend to be shallower in more massive galaxies,
  but this trend is almost entirely the consequence of the fact
  that more massive galaxies have higher $A_V$.
\item Slopes have a secondary dependence on SFR, such that the
  galaxies away from the star-forming sequence in either
  direction (towards starbursts and towards quiescent galaxies), have
  somewhat steeper slopes than the galaxies of the same mass closer to the
  main sequence.
\item Consequently, the analogs of high-redshift galaxies have, on average,
  somewhat steeper curves than the normal star-forming galaxies of the
  same mass, i.e., similar to or steeper than the SMC curve.
\item Galaxies exhibit a range of bump strengths, but rarely exceeding
  the MW value. Stronger bumps tend to be found in galaxies with
  steeper curves, as previously found at higher redshifts
  \citep{kriek13}, but the correlation is not tight. On average, the
  contribution of the bump to near-UV attenuation is relatively small
  ($\sim 1/3$ of that of the MW bump), and can be to first order
  ignored, as it affects the NUV magnitude by only 0.1 mag.
\item Neither the slopes nor the bump strengths have a strong
  dependence on gas metallicity.
\item Galaxies above $\log M_*=10$ have attenuation curve slopes that
  exhibit a moderate dependence on galaxy inclination for $b/a>0.6$,
  which, as in the case of mass dependence, is fundamentally due to
  the dependence of the attenuation curve slope on optical opacity.
\end{enumerate}

We also present functional fits for dust attenuation curves suitable
for use in low and high-redshift studies.  The current work has
significant implications for the study of IRX--$\beta$ and
$A_{\rm FUV}$--$\beta$ relations (e.g., \citealt{boquien09,boquien12})
and will be a subject of a forthcoming publication.

The catalog of lR luminosity-constrained SED fitting parameters used
in this work, such as the stellar mass, dust attenuation and the SFR
are publicly released as GSWLC-2.

\acknowledgments The construction of GSWLC was funded through NASA
ADAP award NNX12AE06G. We thank Daniella Calzetti and Veronique Buat
for valuable discussions. Funding for SDSS-III has been provided by
the Alfred P. Sloan Foundation, the Participating Institutions, the
National Science Foundation, and the U.S. Department of Energy Office
of Science. The SDSS-III web site is http://www.sdss3.org/. Based on
observations made with the NASA Galaxy Evolution Explorer. GALEX is
operated for NASA by the California Institute of Technology under NASA
contract NAS5-98034. This publication makes use of data products from
the Wide-field Infrared Survey Explorer, which is a joint project of
the University of California, Los Angeles, and the Jet Propulsion
Laboratory/California Institute of Technology, funded by the National
Aeronautics and Space Administration.

\section*{Appendix}
\subsection*{A.1. Mock SED fitting} 

\begin{figure*}
\epsscale{1.15} \plotone{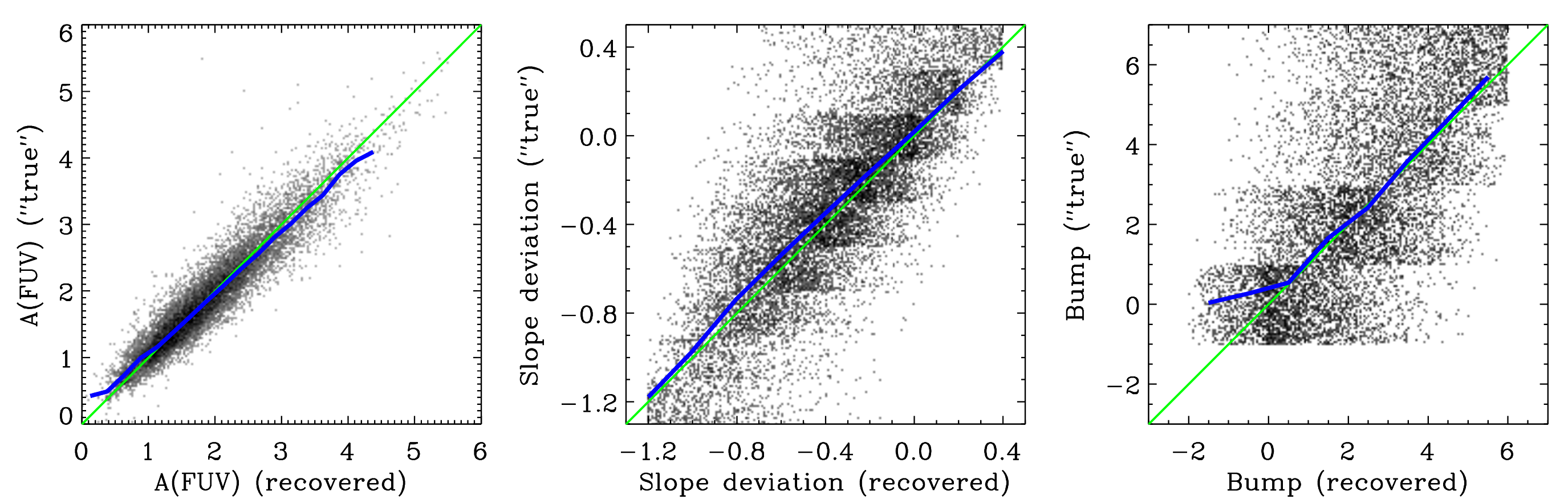}
\caption{Test for systematics in SED fitting parameters using mock SED
  fitting. Known (``true'') values of the effective FUV attenuation
($A_{\rm FUV}$), slope deviation ($\delta$), and UV bump amplitude are
compared with recovered values of these parameters, assuming the
actual photometry (and IR luminosity) errors. Blue line is the running
median, while the green line is a 1:1 relation. Dispersion is
indicative of the parameter errors.
($B$)\label{fig:mock}}
\end{figure*}

In oder to assess whether the choice of priors and the potentially
limited constraining power of the data lead to significant systematics
in the derived values of the attenuation curve slope and UV bump
amplitude, we perform, using CIGALE, mock SED fitting in which we take
the photometry of the best fitting model from the nominal run and
repeat the SED fitting with this photometry instead of the actual one,
while using the actual photometry errors \citep{s09}. The model
photometry, which is treated as input photometry in mock fitting, has
associated ``true'' physical parameters that are known. Mock fitting
allows us to see how well are these parameters recovered. The results
are shown in Figure \ref{fig:mock}, for effective FUV attenuation
($A_{\rm FUV}$), slope deviation ($\delta$), and UV bump amplitude
($B$). True values of slope and the bump are highly discrete (0.2 and
2), so we perturb them by a random value to show the spread. For the
bump amplitude we show only the points with physical true bump
amplitude ($B>0$). Notably, there are no significant systematics in
the recovered values.

\end{document}